\documentclass[twocolumn,prstab,floatfix]{revtex4}
\usepackage{graphicx}
\usepackage{latexsym}
\usepackage{amsfonts}
\usepackage{amssymb}
\usepackage{bm}
\usepackage{subfigure} 
\usepackage[intlimits]{amsmath}

\newcommand{\BfE}{{\bf E}}

\newcommand{\BfJ}{{\bf J}}
\newcommand{\Bfx}{{\bf x}}
\newcommand{\BfX}{{\bf X}}
\newcommand{\Bfu}{{\bf u}}
\newcommand{\Bfr}{{\bf r}}
\newcommand{\stilde}{\widetilde{s}}

\newcommand{\Sin}[1]{\sin \left( #1 \right)}
\newcommand{\Cos}[1]{\cos \left( #1 \right)}

\newcommand{\Exp}[1]{\exp \left( #1 \right)}
\newcommand{\dd}{\; \mathrm{d}}
\DeclareMathOperator{\sgn}{sgn}

\newcommand{\calE}{{\cal E}}
\newcommand{\Csr}{{\mbox{\tiny CSR}}}
\newcommand{\LW}{{\mbox{\tiny LW}}}
\newcommand{\Sc}{{\mbox{\tiny SC}}}

\begin{document}

\title{Exact 1-D Model for Coherent Synchrotron Radiation with Shielding and Bunch Compression}

\author{Christopher Mayes, Georg Hoffstaetter}
\affiliation{Cornell University, Ithaca, New York 14853}

\begin{abstract}
Coherent Synchrotron Radiation has been studied effectively using a 1-dimensional model for the charge distribution in the realm of small angle approximations and high energies. Here we use Jefimenko's form of Maxwell's equations, without such approximations, to calculate the exact wake-fields due to this effect in multiple bends and drifts. It has been shown before that the influence of a drift can propagate well into a subsequent bend. We show, for reasonable parameters, that the influence of a previous bend can also propagate well into a subsequent bend, and that this is especially important at the beginning of a bend. Shielding by conducting parallel plates is simulated using the image charge method. We extend the formalism to situations with compressing and decompressing distributions, and conclude that simpler approximations to bunch compression usually overestimates the effect. Additionally, an exact formula for the coherent power radiated by a Gaussian bunch is derived by considering the coherent synchrotron radiation spectrum, and is used to check the accuracy of wake-field calculations.
\end{abstract}

\maketitle

\section{Introduction}

Coherent Synchrotron Radiation (CSR) is an important detrimental effect in modern particle accelerators with high bunch charges and short bunch lengths. It is a collective phenomenon where the energy radiated at wavelengths on the order of the bunch length is enhanced by the number of charges in the bunch. This CSR field can subsequently affect particle motion. A comprehensive collection of papers and references regarding this subject is found in Ref.~\cite{murphy04}.

CSR is difficult to model using discrete particles exactly because the problem scales with the number of particles $N$ as $N^2$. To simplify this, the 1-dimensional model projects the transverse particle density onto the longitudinal dimension. Formulas for the CSR field from this line charge are then used to calculate forces on each particle and then propagate the full bunch distribution.  While this makes the calculation tractable, the electromagnetic fields on the world-sheet of this charged line are singular. Pioneering efforts described in Refs.~\cite{derbenev95,murphy95} circumvent this problem by examining the non-singular terms only. In more detail, Ref.~\cite{saldin97} regularizes the longitudinal force between two charges traveling on the arc of a circle by subtracting off the Coulomb force, calculated as if the charges traveled on a straight line, from the force calculated using Li\'{e}nard-Wiechert fields. The result is an always finite CSR force.

This paper uses the less widely known Jefimenko forms of Maxwell's equations \cite{jackson99}, which allow one to calculate electromagnetic fields by directly using the evolving charge and current densities, and which internally incorporate all retardation effects. These equations are related to forms used in Refs.~\cite{derbenev95,warnock05}. This is in contrast to the usual Li\'{e}nard-Wiechert approach, which gives fields due to charges at their retarded times $t'$ and positions $\Bfx(t')$, and one must invert equations of the form $t-t' =  |\Bfx(t')-\Bfx_o|/c$ for the retarded time $t'$, where $\Bfx_o$ is an observation point at a later time $t$ and $c$ is the speed of light. While this latter method has proven useful in deriving equations for (incoherent) synchrotron radiation of single particles, the former is found to be useful for the coherent fields of particle distributions.

\section{Exact 1D approaches to steady state CSR}

In general, for given charge and current densities $\rho(\Bfx, t)$ and $\BfJ(\Bfx, t)$ at position $\Bfx$ and time $t$, the electric field $\BfE(\Bfx, t)$ can be calculated using Jefimenko's form of Maxwell's equations \cite{jackson99}
\begin{equation}
\begin{split}
\BfE(\Bfx, t) =  \frac{1}{4\pi \epsilon_0} \int\! \!\!\dd^3 x' &\left[ \frac{\Bfr}{r^3}\, \rho(\Bfx', t')+\! \frac{\Bfr}{c\, r^2}\, \partial_{t'} \rho(\Bfx', t')\right.\\
&\left. -\frac{1}{c^2\, r} \, \partial_{t'} \BfJ(\Bfx', t')\right]_{t' = t-r/c},
 \label{eq:jefimenko}
\end{split}
\end{equation}
in which $\Bfr \equiv \Bfx - \Bfx'$, $r \equiv \|\Bfr\|$,  $\epsilon_0$ is the vacuum permittivity and $t'$ is the retarded time. In this formulation, the retarded points $\Bfx'$ and times $t'$ are independent variables, so there are no functions that need to be inverted.  Therefore, if one knows $\rho$, $\dot{\rho}$, and $\dot{\BfJ}$ at all points in space $\Bfx'$ and times $t'\le t$, with a dot denoting the time derivative, then this formula gives the electric field by direct integration.

Now consider a line charge distribution, which follows a path $\BfX(s)$ parameterized by distance $s$, has a unit tangent $\Bfu(s)$ = $d \BfX(s) / ds$, and  moves with constant speed $\beta c$ along this path. A bunch with total charge $Q$ and normalized line density $\lambda$ therefore has one-dimensional charge density and current
\begin{equation}
\begin{split}
\rho(s, t) &= Q\,  \lambda(s -s_b  -\beta \,c t ),  \\ 
\BfJ(s,t)  &=  Q\,  \beta c \, \Bfu(s) \, \lambda(s -s_b -  \beta \,c t ),
\end{split}
\label{eq:rhosteady}
\end{equation}
where $s_b$ is the location of the bunch center at time $t=0$.

The rate of energy change per unit length of an elementary charge $q$ at position $s$ is $d \calE /ds = q \, \Bfu(s) \cdot \BfE(s,t)$. Functions of this type are called wake-fields. Using Eq.~\eqref{eq:jefimenko} with the one-dimensional bunch in Eq.~\eqref{eq:rhosteady} gives
\begin{equation}\label{eq:1dwake}
\begin{split}
&\frac{d\calE}{ds}(s,t) =N r_c m c^2 \int_{-\infty}^{\infty}\!\!\! \dd  s' \left[  \frac{\Bfu(s) \cdot \Bfr(s,s')}{r(s,s')^3}\lambda(s_r) \right.\\
&\left. - \beta\frac{ \Bfu(s) \cdot \Bfr(s,s')}{r(s,s')^2}\lambda'(s_r)+\beta^2\frac{ \Bfu(s) \cdot \Bfu(s')}{r(s,s')} \lambda'(s_r)\right],\\
\end{split}
\end{equation}
with the definitions
\begin{flalign}
s_r &\equiv s' -s_0 + \beta \, r(s,s'), \\
s_0 &\equiv s_b+\beta\,c t,\\
\Bfr(s,s') &\equiv \BfX(s) - \BfX(s'),\\
r(s,s') &\equiv \|\Bfr(s,s') \|,
\end{flalign}
where $N = Q/q$ is the number of elementary particles with mass $m$ and classical radius $r_c = q^2 /\left( 4 \pi \epsilon_0 m c^2\right)$, and the prime on $\lambda$ indicates a derivative of this function with respect to its argument, \emph{i.e.}\ $\lambda'(x)=d\lambda/dx$. Additionally, $s_0$ is the center of the bunch at time $t$, and this is the only place where the time dependence appears. The integrand is thus the contribution to the wake-field due to particles between the retarded positions $s'$ and $s'+\dd s'$, with $N\,\lambda(s_r)$ being the charge density at retarded position $s'$ and retarded time $t'$.

Unfortunately the integral in  Eq.~\eqref{eq:1dwake} diverges as $s-s' \rightarrow 0$, which is a consequence of the one-dimensional line charge model. This problem can be alleviated by  using the regularization procedure originating in Saldin \emph{et al.}\ \cite{saldin97}, where the electric field $\BfE$ is split into two parts
\begin{equation}
\BfE = \BfE_\Csr + \BfE_\Sc.
\label{eq:Edef}
\end{equation}
The space charge (SC) part is the electric field of a line charge moving on a straight path,
\begin{equation}
\begin{split}
\BfE_\Sc(s,t) =  & \frac{Q}{4\pi \epsilon_0}  \Bfu(s) \int_{-\infty}^{\infty} \dd \stilde \left[\frac{s-\stilde}{|s-\stilde|^3}\lambda(s_l) \right. \\
&  \left. - \beta\frac{s-\stilde}{|s-\stilde|^2 }\lambda'(s_l)+\beta^2\frac{1}{|s-\stilde |} \lambda'(s_l)\right],
\label{eq:sclong}
\end{split}
\end{equation}
with $s_l\equiv\stilde -s_0+ \beta \, |s-\stilde |$, which can be integrated by parts, simplifying to
\begin{equation}
\BfE_\Sc(s,t) = - \frac{Q\,\Bfu(s)}{4\pi \epsilon_0 \gamma^2}  \int_{-\infty}^{\infty}\! \dd \stilde\, \frac{\lambda'\left( \stilde -s_0 + \beta  |s-\stilde |\right)}{|s-\stilde|}  .
\label{eq:ESC}
\end{equation}
It will turn out to be useful to change variables in this expression, so that when combined with Eq.~\eqref{eq:1dwake} the function $\lambda'$ can be factored. This can be done by setting $\stilde -s_0+ \beta \, |s-\stilde | = s' -s_0 + \beta \, r(s,s')$, with the convention that $\sgn(s-\stilde) = \sgn(s-s')$. Noting that  $ \partial r(s,s')/ \partial s' = - \Bfr(s,s') \cdot \Bfu(s') / r(s,s')$, this leads to
\begin{flalign}
\frac{-1}{|s-\stilde|} &= \sgn(s'-s)\frac{1+\beta\,\sgn(s'-s)}{s-s'-\beta\,r(s,s')},\\
\dd \stilde &= \frac{1-\beta\,\Bfr(s,s') \cdot \Bfu(s') / r(s,s')}{1+\beta\,\sgn(s'-s)}\dd s',
\end{flalign}
so that
\begin{equation}\label{eq:ESC2}
\begin{split}
\BfE_\Sc(s,t) =&  \frac{Q}{4\pi \epsilon_0} \frac{ \Bfu(s)}{\gamma^2}  \int_{-\infty}^{\infty} \dd s' \sgn(s'-s) \lambda'(s_r)\\
 &\times\frac{1-\beta\, \Bfr(s,s') \cdot \Bfu(s')/r(s,s')}{s-s'-\beta r(s,s')}.
\end{split}
\end{equation}
The resulting wake-field due to $\BfE_\Csr$, called the CSR-wake, is
\begin{equation}
\left(\frac{d \calE_{\Csr}}{d s}\right)   = q \, \Bfu (s) \cdot  \left[\BfE(s,t) - \BfE_\Sc(s,t) \right].
\label{eq:ECSRdef}
\end{equation}
This expression is finite, and shown in Ref.~\cite{saldin97} to correctly account for the coherent energy loss due to synchrotron radiation.

The approach here is to be contrasted with the conventional one taken in the literature using Li\'{e}nard-Wiechert formulas. In terms of the quantities above, the electric field at position $s$ due to a charge $q$ at retarded time $t'  = t - r(s,s')/c$ and retarded position $s' $ is
\begin{equation}\label{eq:LW}
\begin{split}
\BfE_\LW(s,s') =& \frac{q}{4 \pi \epsilon_0}\left\{\frac{\Bfr - \beta \, r \, \Bfu(s')  }{\gamma^{2}\left[r -\beta \, \Bfr \cdot \Bfu(s')\right]^3}\right. \\
&+\left. \frac{\Bfr \times \left \{ \left[\Bfr - \beta \, r \,  \Bfu(s')\right]\times \beta^2\, \Bfu'(s') \right \}}{\left[r -\beta \, \Bfr \cdot \Bfu(s')\right]^3}\right\},
\end{split}
\end{equation}
with $\Bfr$ as in Eq.~\eqref{eq:1dwake} suppressing the arguments. Therefore, the electric field at $s$ due to a charge $\rho(s_t, t) \dd s_t$ between $s_t$ and  $s_t+\!\dd s_t$, as in Eq.~\eqref{eq:rhosteady}, is found by inverting $s_t = s' + \beta\,  r(s,s') $ for $s'$ and using Eq.~\eqref{eq:LW}. This is often impossible to do analytically, but fortunately for a distribution of charges the inversion can be circumvented by changing variables. Because $ \partial r(s,s')/ \partial s' = - \Bfr \cdot \Bfu(s') / r$ from before, the charge is
\begin{equation}\label{eq:rhoelement}
\begin{split}
\rho(s_t, t) \dd s_t = & Q\,  \lambda (s' - s_b - \beta \, c t + \beta \, r) \\
 & \times\left[1-\beta \frac{\Bfr \cdot \Bfu(s')}{r} \right] \dd s',
\end{split}
\end{equation}
and the total electric field is
\begin{flalign}
&\BfE(s,t)  = \int_{-\infty}^{\infty}\!\! \dd s_t\: \BfE_\LW \!\left(s, s'(s_t)\right)\, \rho (s_t,t) \\
&\qquad=Q \! \int_{-\infty}^{\infty}\!\!\dd s' \left[1-\beta \frac{\Bfr \cdot \Bfu(s')}{r} \right] \BfE_\LW(s, s')  \, \lambda(s_r)  .
\end{flalign}

We can use this to verify that $\BfE_\Sc$ computed this way agrees with the result using the Jefimenko approach. For the SC field one has $\Bfr(s) = (s-s') \Bfu(s)$,  $\Bfu(s') = \Bfu(s)$, and $\Bfu'(s) = 0$, giving
\begin{equation}\label{eq:ESCLW}
\begin{split}
\BfE_\Sc(s,t) =  \frac{Q\,\Bfu(s)}{4\pi \epsilon_0}  \int_{-\infty}^{\infty}\!\! &\dd s' \: \frac{\beta + \sgn(s-s')}{(s-s')^2} \\
& \times \lambda\left( s' -s_0 + \beta \, |s-s' |  \right).
\end{split}
\end{equation}
Equation~\eqref{eq:ESCLW} agrees with Eq.~\eqref{eq:ESC} when integrated by parts because, for $s_0=0$, $\int(\beta + \sgn(s-s'))(s-s')^{-2} \dd s' = (\beta + \sgn(s-s'))(s-s')^{-1}$, and $-\frac{\partial}{\partial s'}\lambda( s' + \beta \,(s-s')\sgn(s-s')) = -(1-\beta \sgn(s-s'))\lambda'( s' + \beta \,|s-s'|)$, and similarly for all $s_0$.

\section{Single Bending Magnet}
\begin{figure}[t]
\centering
\includegraphics[width=0.25\textwidth]{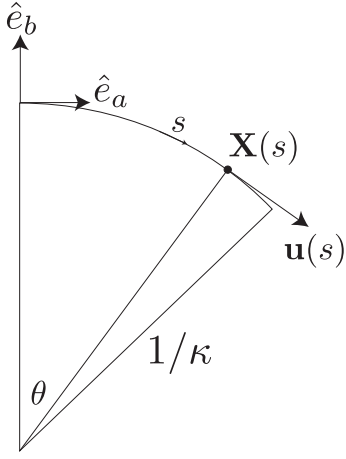}
\caption{Geometry for a single bend. The variable $s$ parameterizes the curve with radius $1/\kappa$. The coordinates are $\BfX(s)$, and the unit tangent vector is $\Bfu(s)$.}
\label{fg:bendgeometry}
\end{figure}
Now we apply Eq.~\eqref{eq:1dwake} to the geometry of an arc of a circle of curvature $\kappa$ and length $B$, shown in Fig.~\ref{fg:bendgeometry}. Set $s=0$ at the entrance of the bend so that $\theta = \kappa \, s$ is the angle into the bend. In terms of fixed Cartesian unit vectors $\hat{e}_a$ and $\hat{e}_b$, the path coordinates and tangent vector are
\begin{flalign}
\BfX(s) &= \kappa^{-1} \Sin{ \kappa \, s }\hat{e}_a - \kappa^{-1} \left[1- \Cos {\kappa \, s} \right] \hat{e}_b, \\
\Bfu(s) &= \Cos {\kappa \, s}  \hat{e}_a - \Sin{ \kappa \, s} \hat{e}_b.
\end{flalign}

Consider a bunch with its center at angle $\theta_0 = \kappa\,  s_0$, and a test particle at angle $\theta$. The contribution to Eq.~\eqref{eq:1dwake} of this finite arc is
\begin{equation}\begin{split}
\left.\frac{d\calE}{ds}(s)\right|_B  &=N r_c m c^2 \int_{0}^{\kappa \,  B}  \dd \theta' \left[\frac{\sin \alpha }{(\kappa\, r_\alpha)^3}\kappa\,  \lambda(s_\alpha)\right. \\
&\left.- \beta\frac{ \sin \alpha}{(\kappa\, r_\alpha)^2}\lambda'(s_\alpha)+\beta^2\frac{\cos \alpha}{\kappa\, r_\alpha} \lambda'(s_\alpha)\right], \label{eq:intbend}
\end{split}\end{equation}
with $s=\kappa^{-1}\theta$ and the following definitions:
\begin{flalign}
\alpha &\equiv \theta - \theta',\\
s_\alpha &\equiv \frac{1}{\kappa} \left( \theta  - \theta_0 - \alpha\right)+ \beta \, r_\alpha, \label{eq:salpha}\\
r_\alpha &\equiv \frac{1}{\kappa}\sqrt{2-2 \cos \alpha}.
\end{flalign}
Thus $\alpha$ is the angle between the test particle and the retarded source particle, and is positive when the former is ahead of the latter. The first term of Eq.~\eqref{eq:intbend} can be integrated by parts because $\partial (2-2\cos \alpha)^{-1/2}  / \partial \theta' = \Sin{\alpha} (2-2\cos \alpha)^{-3/2}$, and the wake greatly simplifies to
\begin{equation}
\begin{split}
\left.\frac{d\calE}{ds}(s)\right|_B =& N r_c m c^2\Bigg \{\left. \frac{-\kappa \, \lambda(s_\alpha)}{\sqrt{2-2 \cos \alpha}}\right|_{\alpha=-( \kappa \, B - \theta)}^{\alpha=\theta}  \\
& +\int_{-( \kappa \, B - \theta)}^{\theta} \!\!\!\!\!\!\dd \alpha \: \frac{\beta^2 \Cos{ \alpha}-1}{\sqrt{2-2 \cos \alpha} } \lambda' (s_\alpha) \Bigg\}.
\label{eq:intbend2}
\end{split}
\end{equation}
In terms of the variable $\alpha$, the space charge term in Eq.~\eqref{eq:ESC2} can be split as
\begin{equation}
\begin{split}
\frac{d\calE_\Sc}{ds}(s) =& -N r_c m c^2 \left\{  \int_{-\infty}^{-(\kappa \, B - \theta)}\!\!\!\!\!\! \dd \alpha \: I_{\Sc}(\alpha) \right. \\
&\left. + \int_{-(\kappa \, B - \theta)}^{\theta}\!\!\!\!\!\! \dd \alpha \: I_{\Sc}(\alpha)
+ \int_{ \theta}^{\infty} \dd \alpha \: I_{\Sc}(\alpha)
\right \},
\end{split}
\end{equation}
with the integrand
\begin{equation}
 I_{\Sc} (\alpha) \equiv -\frac{\sgn \alpha}{\gamma^2}\frac{1- \dfrac{\beta\,\Sin{\alpha}}{\sqrt{2-2\cos \alpha}}}{ \alpha-\beta \, \sqrt{2-2\cos \alpha}}\,\lambda'( s_\alpha ),
 \label{eq:ISCdef} %
\end{equation}
so that the contribution of the bend to the CSR-wake is
\begin{equation}
\begin{split}
&\left.\frac{d\calE_{\Csr}}{ds}(s)\right|_B  =N r_c m c^2  \Bigg\{ \!\!\left.  \frac{-\kappa \, \lambda(s_\alpha)}{\sqrt{2-2 \cos \alpha}}\right|_{\alpha=-( \kappa \, B - \theta)}^{\alpha=\theta} \\
 &\qquad+ \!\!\int_{-( \kappa \, B - \theta)}^{\theta}\!\!\!\!\!\!\! \dd \alpha \:   \lambda' (s_\alpha)\Bigg[\frac{\beta^2 \Cos{ \alpha}-1}{\sqrt{2-2\cos \alpha}} \\
 &\qquad\qquad\qquad\qquad+ \frac{\sgn (\alpha)}{\gamma^2}\frac{1- \dfrac{\beta\,\Sin{\alpha}}{\sqrt{2-2\cos \alpha}}}{ \alpha-\beta \, \sqrt{2-2\cos \alpha}} \Bigg]  \\
 &\qquad - \int_{-\infty}^{-(\kappa \, B - \theta)} \!\!\!\!\!\!\!\dd \alpha \: I_{\Sc}(\alpha) - \int_{\theta}^{\infty} \dd \alpha \: I_{\Sc} (\alpha) \Bigg \}.
\end{split}
\label{eq:ECSRbend}
\end{equation}

\subsection{Steady State}
In the practical environment of a particle accelerator with a bunched beam, one is typically only concerned with electric fields around the bunch center. Due to the rotational symmetry, there will be an angle into a bending magnet beyond which the CSR-wake, relative to the bunch center, does not change. Note that in Eq.~\eqref{eq:salpha} the quantity $z = \kappa^{-1}(\theta-\theta_0)$ is the distance along the path ahead of the bunch center, and define the extent of the bunch $l_b \equiv z_+ - z_{-}$, where $z_+$ is the head particle coordinate, and $z_-$ is the tail particle coordinate. Henceforth the symbol $z$ will refer to the longitudinal coordinate relative to the bunch center: $z=s-s_0$. The particle at $z_+$ is affected by a particle at $z_-$ at retarded angle $\alpha_{\rm max} $ found by inverting
\begin{equation}
\kappa \, l_b = \alpha_{\rm max}  - \beta \sqrt{2-2\cos \alpha_{\rm max} }.
\end{equation}
Similarly, a particle at $z_-$ is affected by a particle at $z_+$ at retarded angle $\alpha_{\rm min}$ found by inverting
\begin{equation}
-\kappa \, l_b = \alpha_{\rm min}  - \beta \sqrt{2-2\cos \alpha_{\rm min} }.
\end{equation}

When the bunch center is at an angle $\theta_0 > \alpha_{\rm max} - \kappa \, z_+$, only particles within the bend affect the wake-field. The ``steady-state'' (s.s.) CSR-wake is then
\begin{equation}
\begin{split}
&W_{\begin{subarray}{|}\Csr \\ {\rm s.s.}\end{subarray}} (z) =N r_c m c^2   \int_{\alpha_{\rm min}}^{\alpha_{\rm max}}\!\! \dd \alpha \:  \Bigg[\frac{\beta^2 \Cos{ \alpha}-1}{\sqrt{2-2\cos \alpha}} \\
&\qquad  + \frac{\sgn (\alpha)}{\gamma^2}\frac{1-\dfrac{\beta\, \Sin{\alpha}}{\sqrt{2-2\cos \alpha}}}{ \alpha-\beta \, \sqrt{2-2\cos \alpha}} \Bigg]\lambda' \left(z-\Delta(\alpha)\right) ,
\label{eq:Esteadystate}%
\end{split}
\end{equation}
where
\begin{equation}
\Delta(\alpha) = \kappa^{-1}(\alpha - \beta \, \sqrt{2-2\cos \alpha})
\label{eq:Delta}
\end{equation}
is the distance behind the test particle at $z$. The notation
\begin{equation}
W_\Csr(z) \equiv \frac{d\calE_{\Csr}}{ds}(s_0+z)
\end{equation}
is used to refer to the CSR-wake immediately surrounding the bunch center at $s_0$.
\begin{figure}[tb!]
\centering
\includegraphics[width=0.45\textwidth]{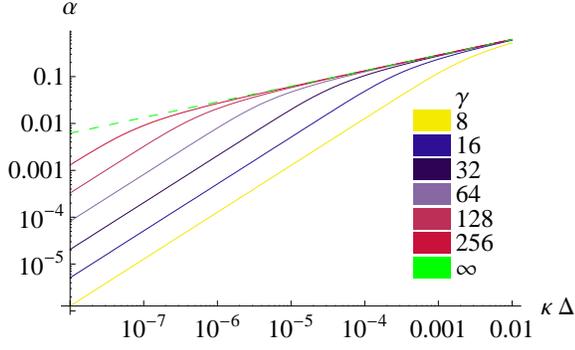}
\caption{The inverse of Eq.~\eqref{eq:Delta} for positive $\alpha$ at various energies. The dashed green curve ($\gamma\rightarrow\infty$) is $\alpha = (24\,\kappa\,\Delta)^{1/3}$, the inverse of Eq.~\eqref{eq:deltaapprox}. }
\label{fg:alpha}
\end{figure}

In the ultra-relativistic approximation ($\beta \rightarrow 1$) with a small normalized bunch length $\kappa \, l_b \ll 1$, and thus $\alpha \ll 1$, the steady-state formula in Eq.~\eqref{eq:Esteadystate} greatly simplifies. The $1/\gamma^2$ term in Eq.~\eqref{eq:ISCdef} puts $I_\Sc \rightarrow 0$, and the term in the integrand $(\beta^2 \Cos{ \alpha}-1)/(2 \left | \Sin{\alpha/2} \right |) \approx -|\alpha|/2$. The function $\Delta(\alpha)$  for $\gamma\rightarrow\infty$ is approximately
\begin{equation}
\Delta(\alpha)\approx
\begin{cases}
\alpha^3 / (24 \kappa) & \text{for $\alpha > 0$}\\
2 \alpha/\kappa & \text{for $\alpha < 0$}\
\end{cases}.
\label{eq:deltaapprox}
\end{equation}
\begin{figure}[tb!]
\centering
\includegraphics[width=0.45\textwidth]{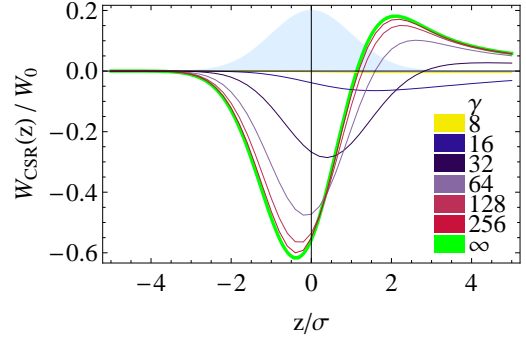}
\caption{The steady-state CSR-wake for various relativistic $\gamma$ using Eq.~\eqref{eq:Esteadystate}, compared to Eq.~\eqref{eq:Esteadystateinfinity} plotted as green. Here $\kappa \, \sigma = 3\times 10^{-5} $ for a Gaussian bunch, represented in light blue.}
\label{fg:steadystatewake}
\end{figure}
\begin{figure}[h]
\centering
\includegraphics[width=0.45\textwidth]{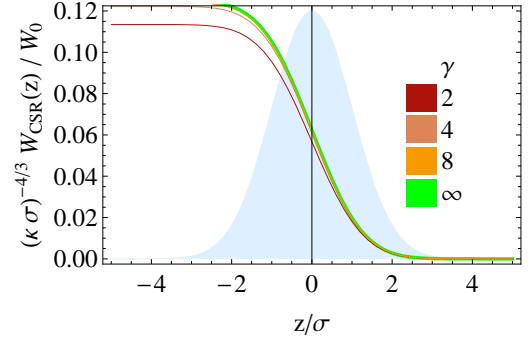}
\caption{The steady-state CSR-wake for various relativistic $\gamma$ due only to particles ahead of the test particle, \emph{i.e.}\ negative $\alpha$ in Eq.~\eqref{eq:Esteadystate}, and the second term in the integrand of Eq.~\eqref{eq:Esteadystateinfinity}. A Gaussian bunch is used, with $\kappa \, \sigma = 3\times 10^{-5} $, and the wake has been scaled by $(\kappa \, \sigma)^{-4/3}$. Compared with Fig.~\ref{fg:steadystatewake} this demonstrates that the contribution to the CSR-wake of particles ahead of the test particle is insignificant compared to those behind.}
\label{fg:steadystatewakeforward}
\end{figure}

\begin{figure}[t!]
\centering
\includegraphics[width=0.45\textwidth]{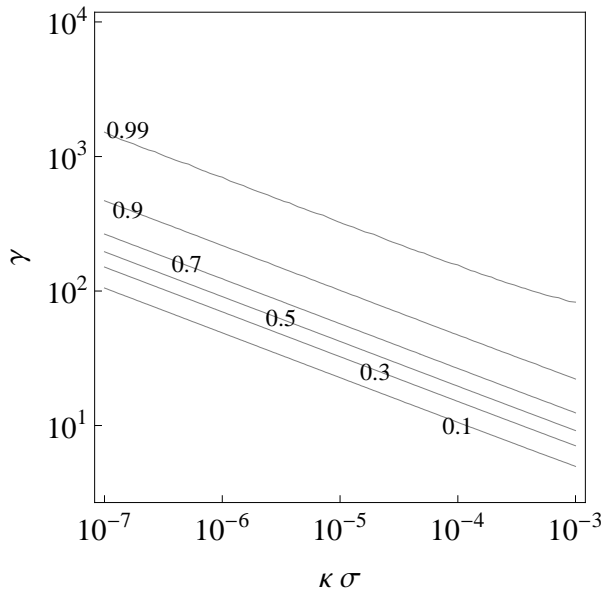}
\caption{The ratio of the average energy of a Gaussian bunch using the exact Eq.~\eqref{eq:Esteadystate} to that using approximate Eq.~\eqref{eq:Esteadystateinfinity} in a practical range of the parameters $\gamma$ and $\kappa\,\sigma$. }
\label{fg:steadystatecomparison}
\end{figure}

Figure~\ref{fg:alpha} plots the inverse of Eq.~\eqref{eq:Delta} for positive $\alpha$ and various energies. One sees that the approximation in Eq.~\eqref{eq:deltaapprox} is increasingly good for higher energies, but greatly overestimates $\alpha$ at the smallest distances. Changing variables using Eq.~\eqref{eq:deltaapprox}, the ultra-relativistic steady-state CSR-wake is
\begin{equation}
\begin{split}
W_{\begin{subarray}{|}\Csr \\ \gamma\rightarrow\infty\end{subarray}} (z) =- N r_c m c^2\,\kappa \int_{0}^{l_b}& \!\!\dd \Delta  \:  \Bigg[   \frac{2\,\lambda' (z-\Delta)}{ (3\, \kappa \,  \Delta)^{1/3}} \\
&+ \frac{\kappa\,  \Delta}{8} \lambda' (z+\Delta) \Bigg].
\label{eq:Esteadystateinfinity}
\end{split}\end{equation}
The first term in this integral is derived by an alternate method in Ref.~\cite{saldin97}. The scaling here is apparent by writing the distribution in the normalized form
\begin{gather}
\lambda(z-\Delta) \equiv \frac{1}{\sigma} \widetilde{\lambda}\left( \frac{z-\Delta}{\sigma}\right), \label{eq:lambdatilde} \\
\lambda'(z-\Delta) \equiv \frac{1}{\sigma^2} \widetilde{\lambda}'\left( \frac{z-\Delta}{\sigma}\right).
\end{gather}
where $\sigma^2$ is the variance of $\lambda$, so that $\widetilde{\lambda}$ has unit variance. Also using normalized $\widetilde{z} \equiv z/\sigma$ and $\widetilde{\Delta} \equiv \Delta/\sigma$ gives
\begin{equation}
\begin{split}
&W_{\begin{subarray}{|}\Csr \\ \gamma\rightarrow\infty\end{subarray}}(\widetilde{z}\, \sigma) =-N r_c m c^2 \frac{(\kappa\, \sigma)^{2/3}}{\sigma^2}\\
&\quad\times \int_{0}^{l_b/\sigma}\!\!\! \dd \widetilde{\Delta}
 \Bigg[   \frac{2\,\widetilde{\lambda}' (\widetilde{z}-\widetilde{\Delta})}{ (3\, \widetilde{\Delta})^{1/3}}
+ (\kappa \, \sigma)^{4/3}\: \frac{  \widetilde{\Delta}}{8} \widetilde{\lambda}' (\widetilde{z}+\widetilde{\Delta}) \Bigg].
\label{eq:Esteadystateinfinitynormalized}
\end{split}
\end{equation}
Now one can see that the particles in front of the test particle, represented in the last term in the integrand, influence the wake by roughly a factor of $(\kappa \, \sigma)^{4/3}$ less than particles behind, and that the primary contribution to the CSR-wake scales with the factor in front of the integral in Eq.~\eqref{eq:Esteadystateinfinitynormalized}. However, it is interesting to note that even as $\gamma\rightarrow\infty$, where a charge radiates infinitely more power in the forward direction than the backward direction, there is still a finite CSR force from particles ahead of the test particle. In light of the primary scaling, we define a characteristic CSR energy change per unit length as
\begin{equation}
W_0 \equiv N r_c m c^2 \frac{(\kappa\, \sigma)^{2/3}}{\sigma^2}.
\label{eq:W0}
\end{equation}

The ultra-relativistic approximation in Eq.~\eqref{eq:Esteadystateinfinity} is compared to the exact formula Eq.~\eqref{eq:Esteadystate} in Fig.~\ref{fg:steadystatewake} for various energies and a particular value of $\kappa \, \sigma$. One sees that Eq.~\eqref{eq:Esteadystateinfinity} represents the largest possible effect. The CSR-wake due only to particles in front of the test particle is shown in Fig.~\ref{fg:steadystatewakeforward}, emphasizing again that these forward particles contribute only a small amount to the total CSR-wake.

Neglecting the contribution due to forward particles, the ultra-relativistic stead-state CSR-wake in Eq.~\eqref{eq:Esteadystateinfinitynormalized} scales with $W_0$ and depends only on the shape of $\widetilde{\lambda}$. Factoring out $W_0$ from the exact steady-state CSR-wake in Eq.~\eqref{eq:Esteadystate}, the exact result additionally depends on $\gamma$ and $\kappa\,\sigma$.  Therefore, to quantify the appropriateness of the ultra-relativistic approximation, the ratio of the average energy lost (per unit length) of a Gaussian bunch using the exact Eq.~\eqref{eq:Esteadystate} to that using approximate Eq.~\eqref{eq:Esteadystateinfinity} is shown in Fig.~\ref{fg:steadystatecomparison} for a practical range of these parameters. At a given energy, one sees that Eq.~\eqref{eq:Esteadystateinfinity} is a good approximation for the relatively long bunches. This can be understood from Fig.~\ref{fg:alpha}, because the approximation in Eq.~\eqref{eq:deltaapprox} has a relative error for a finite energy that diverges for small $\alpha$.

\begin{figure}[tb!]
\centering
\includegraphics[width=0.45\textwidth]{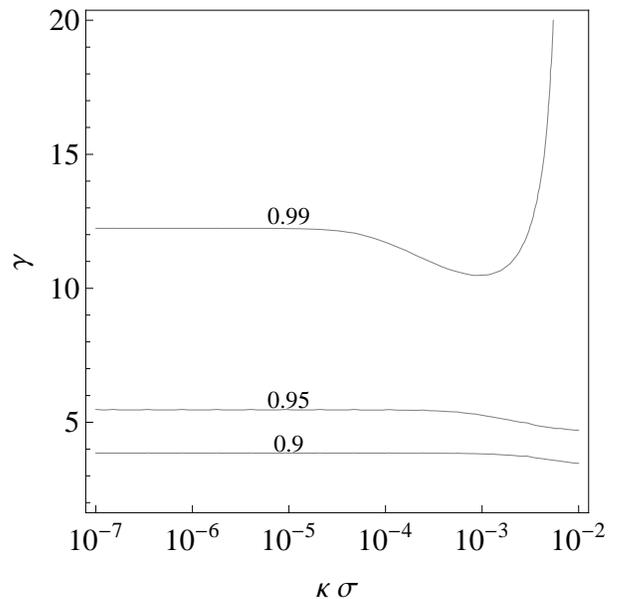}
\caption{Similar to Fig.~\ref{fg:steadystatecomparison}, but with the ratio of the average energy of Eq.~\eqref{eq:Esteadystate} to Eq.~\eqref{eq:EsteadystateSHMS}, showing that the latter is an excellent approximation at relatively low energies. }
\label{fg:SHMSsteadycomparison}
\end{figure}

A systematic method for calculating the CSR-wake using Li\'{e}nard-Wiechert formulas in the small angle, relativistic approximations has been developed in Ref.~\cite{sagan08} for arbitrary combinations of drifts and bends. Using the corresponding equation in Ref.~\cite{sagan08} for the geometry of a bend, and the appropriate Jacobian factor, the steady-state CSR-wake to second order in $\alpha$ and $1/\gamma$ is
\begin{equation}
\begin{split}
 W_{\begin{subarray}{|}\Csr \\ {\rm SHMS, s.s.}\end{subarray}} (z) &= - N r_c m c^2  \,   \int_{0}^{\alpha_{\rm max}}\!\!\! \dd \alpha \Bigg[\left(\frac{1}{2 \gamma^2}+\frac{\alpha^2}{8} \right)\\
&\times\left( \frac{2 +\gamma^2 \alpha^2}{\alpha+\gamma^2 \alpha^3 /4}-\frac{1}{\alpha/2+\gamma^2 \alpha^3/24}\right)\\
&\times \lambda'\left(z-\kappa^{-1}\left(\frac{\alpha}{2 \gamma^2}+ \frac{\alpha^3}{24} \right) \right)\Bigg].
\label{eq:EsteadystateSHMS}%
\end{split}
\end{equation}
Compared to Eq.~\eqref{eq:Esteadystateinfinity}, this expression is a significantly better approximation of Eq.~\eqref{eq:Esteadystate} for low $\gamma$ and a practical range of $\kappa  \sigma$, shown in Fig.~\ref{fg:SHMSsteadycomparison}.

\subsection{Shielding by Parallel Plates}
\begin{figure}[t!]
\centering
\includegraphics[width=0.45\textwidth]{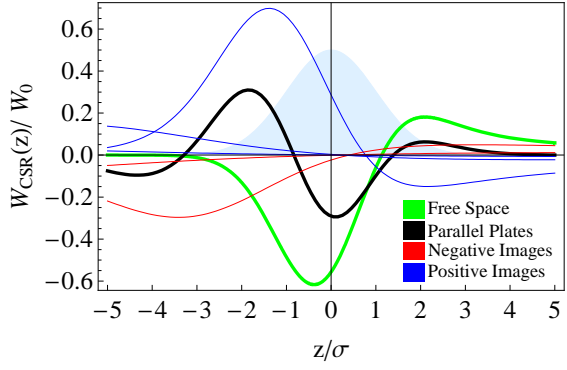}
\caption{The free space and shielded CSR-wakes in a bend. The contributions to the shielded wake of individual image bunches are shown in red and blue. A 1GeV Gaussian bunch with $\sigma=0.3$mm is used in a bend of radius $\kappa^{-1}=10.0$m. The shielding height $H=2$cm.}
\label{fg:imagewakes}
\end{figure}
The presence of a conducting beam chamber can have a strong effect on the CSR wake-field. For a rectangular cross section, it has been observed that the dominant effect comes from the smaller of the height and width (see, for example, Ref.~\cite{sagan08}). If particle trajectories are planar, then a finite chamber height can be represented by infinite parallel plates. In such a geometry, CSR wake-fields can be calculated using the image charge method.

\begin{figure}[t!]
\centering
\includegraphics[width=0.45\textwidth]{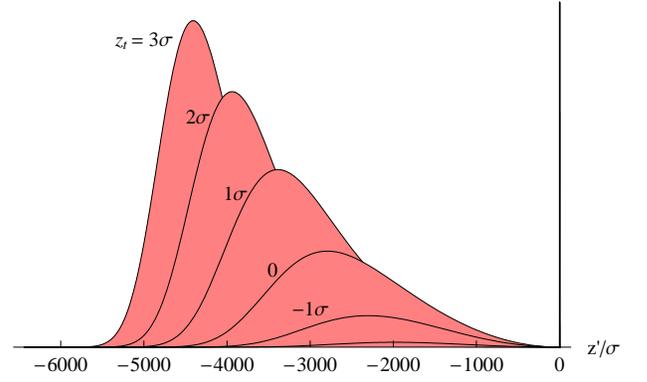}
\caption{The steady-state retarded distribution $\lambda_{\rm ret, s.s.}(z'; z_t)$ for various test particles $z_t$ in Eq.~\eqref{eq:lambdaretss} using a Gaussian bunch with standard deviation $\sigma = 0.3$mm and energy 1GeV, in a magnet of bending radius $\kappa^{-1} = 10.0$m.}
\label{fg:retardedbunches}
\end{figure}

\begin{figure*}[t!]
\centering
\includegraphics[width=\textwidth]{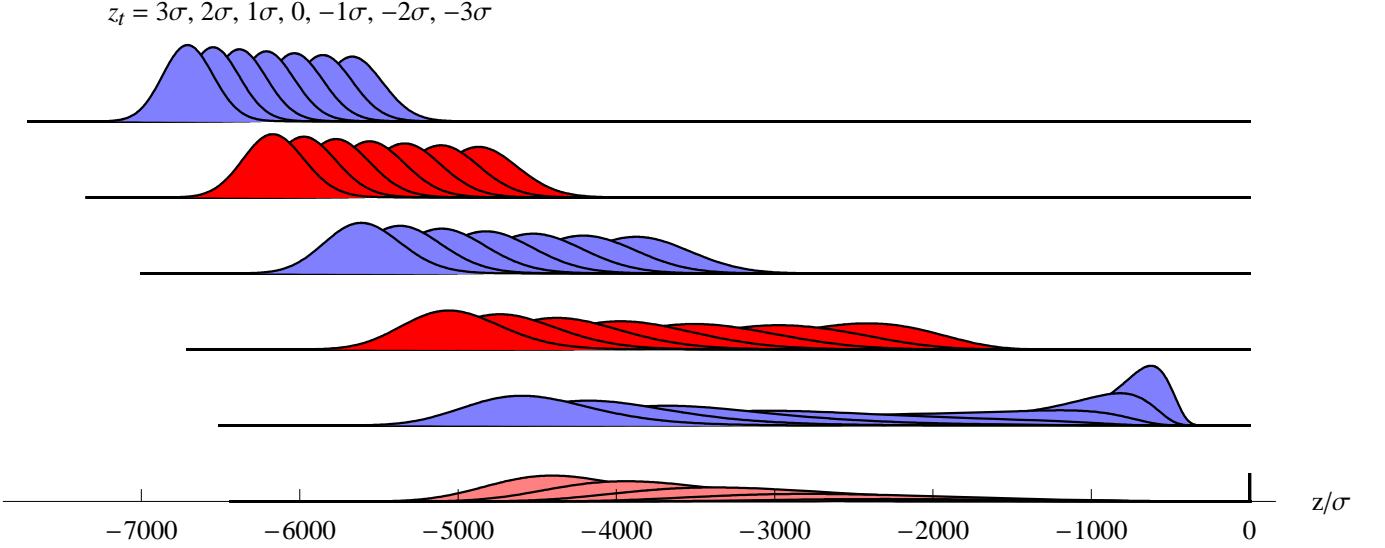}
\caption{The same as Fig.~\ref{fg:retardedbunches}, along with image charges at a heights $n H=n\times2$cm (not to scale), which are approximately at heights $n\times67\sigma$, and calculated using Eq.~\eqref{eq:lambdaretimages}.}
\label{fg:retardedimages}
\end{figure*}

The kick due to a single image bunch at height $h$ is easily adapted from Eq.~\eqref{eq:1dwake} as
\begin{equation}
\begin{split}
&\frac{d\calE}{ds}(s,t,h) =N r_c m c^2 \int_{-\infty}^{\infty}\!\!\!\! \dd  s' \:\Bigg\{  \frac{\Bfu(s) \cdot \Bfr}{(r^2+h^2)^{3/2}}\lambda(s_{h}) \\
&\qquad\quad+ \left[ \beta^2\frac{ \Bfu(s) \cdot \Bfu(s')}{(r^2+h^2)^{1/2}}-\beta\frac{ \Bfu(s) \cdot \Bfr}{r^2+h^2} \right]\lambda'(s_{h})\Bigg\},
 \label{eq:1dwakeimage}
\end{split}
\end{equation}
with the argument
\begin{equation}
s_h \equiv s' -s_b- \beta c \, t + \beta \, \sqrt{r^2+h^2},
\end{equation}
and with $\Bfr$ and $\Bfu$ retaining their meaning from Eq.~\eqref{eq:1dwake}. Parallel plates require an image bunch for each plate, and an image bunch for each of those, \emph{ad infinitum}. For the real bunch with orbit midway between plates separated by a distance $H$, symmetry gives the total image kick
\begin{flalign}
\frac{d\calE_{\rm images}}{ds}(s,t) &= \sum_{\substack{n = -\infty\\ n\neq 0}}^{\infty}(-1)^n \: \frac{d\calE}{ds}(s,t,n\,H)\\
                                    &= 2 \sum_{n = 1}^{\infty} (-1)^n \: \frac{d\calE}{ds}(s,t,n\,H).
\label{eq:imagewake}
\end{flalign}
If the real bunch has a vertical offset $V$, the total image kick is modified to
\begin{equation}
\begin{split}
\frac{d\calE_{\rm images}}{ds}(s,t) = &\sum_{\substack{n =-\infty, n \neq 0 \\ \text{even} }}^{\infty} \frac{d\calE}{ds}(s,t,n\,H) \\
&-\sum_{\substack{n=-\infty \\ \text{odd} }}^{\infty} \frac{d\calE}{ds}(s,t,n\,H - 2V).
\end{split}\end{equation}
In a bend, the contribution of the image bunches to the CSR-wake within the bend, following Eq.~\eqref{eq:intbend2}, is
\begin{equation}
\begin{split}
&\left.\frac{d\calE_{\rm images}}{ds}(s)\right|_B = N r_c m c^2 \\
 &\qquad\times\sum_{n=1}^{\infty} 2(-1)^n \Bigg\{\left . \frac{-\kappa \, \lambda(s_{\alpha,n})}{r_{\alpha, n}}\right|_{\alpha=-( \kappa \, L_m - \theta)}^{\alpha=\theta} \\
 &\qquad \qquad\quad+ \int_{-( \kappa \, L_m - \theta)}^{\theta}\!\!\!\!\!\!\!\!\!\! \dd \alpha \: \frac{\beta^2 \Cos{ \alpha}-1}{r_{\alpha,n} } \lambda'(s_{\alpha, n}) \Bigg\},
\label{eq:intbend2image}
\end{split}
\end{equation}
with the definitions
\begin{flalign}
r_{\alpha, n} &\equiv \sqrt{2-2 \cos \alpha+ (n\, \kappa\,  H)^2}, \\
s_{\alpha, n} &\equiv \kappa^{-1}\left(\theta-\theta_0-\alpha + \beta\, r_{\alpha, n}\right).
\end{flalign}
Notice that the integrands do not need to be regularized by the SC term, because they are always finite due to the always positive factor $(n\, \kappa\, H)^2$.

Due to the infinite number of image layers needed, a finite bend can never be exactly in the steady-state. However, due to their increased distances and angles, the relevant contribution image number $n$ will be negligible beyond some maximum image number. This point is illustrated in Fig.~\ref{fg:imagewakes}, where the contributions to the CSR-wake of five individual images are shown along with their sum with the free space wake, to give the total shielded wake.

\subsection{Retarded Bunch Visualization}

For a given particle at time $t$ within the bunch, it is evident that the retarded bunch density can be very distorted relative to the actual bunch density. From Eq.~\eqref{eq:rhoelement}, the retarded bunch density at position $s'$ as seen by a particle at position $s$ is
\begin{equation}
\lambda_{\rm ret}(s';s) = \lambda (s' - s_b-\beta\,ct + \beta \, r) \left(1-\beta \frac{\Bfr \cdot \Bfu(s')}{r} \right).
\label{eq:lambdaret}
\end{equation}

In the steady-state, the geometry of a bend can be used in Eq.~\eqref{eq:lambdaret}. Moving to coordinates relative to the bunch center, the steady-state density seen by a test particle at $z_t$ within the bunch as a function of $z'$ is
\begin{equation}
\begin{split}
&\lambda_{\substack{{\rm ret}\\{\rm s.s.}}}(z'; z_t) = \left[1-\beta \frac{\Sin{\kappa(z_t-z')}}{\sqrt{2-2\Cos{\kappa(z_t-z')}}} \right] \\
& \qquad\quad \times\lambda\left(z'+\beta \, \kappa^{-1}\sqrt{2-2\Cos{\kappa(z_t-z')}} \right).
\label{eq:lambdaretss}
\end{split}
\end{equation}
This retarded density is illustrated in Fig.~\ref{fg:retardedbunches} for a Gaussian bunch distribution for various test particles. There one sees that the density in front of the test particle is compressed to roughly $\sigma/(1+\beta) \approx \sigma/2$, concentrated in an apparent spike at the right of the plot. The density behind the test particle occupies the majority of the plot. While it may seem that the curves shown are Gaussian in form, this is only true for the left sides of the curves; the right sides have been extended and diluted due to the Jacobian factor in Eq.~\eqref{eq:lambdaret}. Similarly, the retarded density of an image bunch at height $h$ is
\begin{equation}
\begin{split}
\lambda_{\rm ret}(s',h;s) =& \left[1-\beta \frac{\Bfr \cdot \Bfu(s')}{\sqrt{r^2+h^2}} \right]\\
&\times\lambda (s' - s_b-\beta\,ct + \beta \, \sqrt{r^2+h^2}) .
\label{eq:lambdaretimages}
\end{split}\end{equation}

Figure~\ref{fg:retardedimages} shows the retarded densities for a Gaussian bunch and several image bunches within a bend. In this example, the first and second image bunches as seen by particles in the rear of the bunch are actually \emph{closer} than the real retarded bunch.

\section{Multiple Bends and Drifts}
\begin{figure}[t!]
\centering
\includegraphics[width=0.45\textwidth]{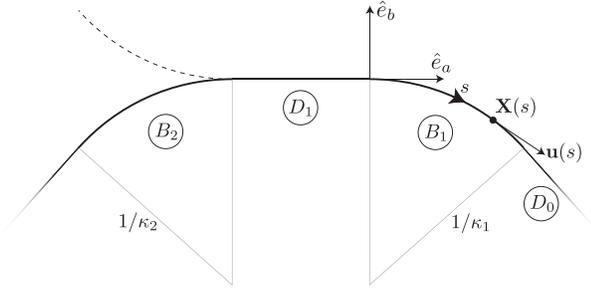}
\caption{Geometry for bends and drifts. The variable $s$ parameterizes the path $\BfX(s)$, with $s=0$ at the beginning of element $B_1$. The names $B_1$, $D_1$, \emph{etc.}, also serve to indicate the element length. The dashed line is for a prior bend with negative curvature.}
\label{fg:twobendgeometry}
\end{figure}

\subsection{CSR in Multiple Bends}

In this section the general formula Eq.~\eqref{eq:1dwake}, regularized by Eq.~\eqref{eq:ESC}, is applied to the geometry of multiply connected bends and drifts. Shielding by conducting parallel plates is added as in Eq.~\eqref{eq:imagewake}. It has been seen in Eq.~\eqref{eq:Esteadystateinfinitynormalized} that the primary contribution to the CSR-wake in a bend is due to particles behind the test particle, so for brevity the path is given behind the test particle only.

Let the bunch center be at length $s_0$ inside bend 1 of length $B_1$ and positive curvature $\kappa_1$, preceded by drift 1 of length $D_1$, preceded by bend 2 of length $B_2$ and curvature $\kappa_2 \neq 0$, as shown in Fig.~\ref{fg:twobendgeometry}. A drift follows bend 1, referred to as $D_0$. A negative curvature $\kappa_2$ signifies a bend in the opposite direction of bend 1. With $s=0$ located at the beginning of bend 1, the path coordinates are
\begin{equation}
\BfX(s) =
\begin{cases}
 \BfX_{D_0}(s)&\text{for $s>B_1$ } \\
\BfX_{B_1}(s) & \text{for $0< s \leq B_1$}\\
s \, \hat{e}_a & \text{for $-D_1 < s \leq 0$ }\\
\BfX_{B_2}(s) & \text{for $s \leq -D_1$}\\
\end{cases}
\label{eq:Xofs}
\end{equation}
where the paths in the individual elements are
\begin{equation}\begin{split}
&\BfX_{D_0}(s)\equiv \left[\frac{\Sin{\kappa_1 B_1}}{\kappa_1}+(s-B_1)\Cos{\kappa_1 B_1} \right]\hat{e}_a   \\
 &+\left[\frac{1}{\kappa_1}(\Cos{\kappa_1 B_1}-1)-(s-B_1) \Sin{\kappa_1 B_1}\right]\hat{e}_b,
\end{split}\end{equation}
\begin{equation}
\BfX_{B_1}(s)\equiv\frac{\Sin{ \kappa_1 \, s }}{\kappa_1}\,\hat{e}_a -\frac{\left[1- \Cos {\kappa_1 \, s} \right]}{\kappa_1}  \,\hat{e}_b,
\end{equation}
\begin{equation}\begin{split}
\BfX_{B_2}(s)\equiv& \left[\frac{\Sin{ \kappa_2 \, D_1+\kappa_2\,s }}{\kappa_2} -D_1\right]\,\hat{e}_a \\
&-\frac{1- \Cos {\kappa_2 \, D_1+ \kappa_2\,s}}{\kappa_2} \, \hat{e}_b.
\end{split}\end{equation}
The tangent vector is then
\begin{equation}
\Bfu(s)=
\begin{cases}
\begin{array}{l}
\Cos{\kappa_1 B_1}\hat{e}_a    \\
 \quad  -\Sin{\kappa_1 B_1}\hat{e}_b
\end{array} &\!\!\!\!\!\!\text{for $s>B_1$ } \\
\begin{array}{l}
\Cos{ \kappa_1 \, s }\hat{e}_a   \\
 \quad -\Sin {\kappa_1 \, s} \hat{e}_b
\end{array} &\!\!\!\!\!\!\text{for $0< s \leq B_1$} \\
1\hat{e}_a & \!\!\!\!\!\!\text{for $-D_1 < s \leq 0$ }\\
\begin{array}{l}
\Cos{ \kappa_2  D_1+\kappa_2s } \hat{e}_a   \\
\: -\Sin {\kappa_2  D_1+\kappa_2s}  \hat{e}_b
\end{array} &\!\!\!\!\!\!\text{for $s \leq -D_1$} \\
\end{cases}
\label{eq:uofs}
\end{equation}
Straightforward calculation gives the total CSR-wake at position $s$ in the bend ($0<s<B_1$) due to these different sections of the path
\begin{equation}
\begin{split}
\left.\frac{d\calE_{\Csr}}{ds}\right|_{\rm tot}& (0<s<B_1) = \left.\frac{d\calE_{\Csr}}{ds}\right|_{B_1} \\ &+\left.\frac{d\calE_{\Csr}}{ds}\right|_{D_1} +\left.\frac{d\calE_{\Csr}}{ds}\right|_{B_2} + \ldots
\label{eq:totalbendwake}
\end{split}
\end{equation}
with $B_1$, $D_1$, and $B_2$ signifying the contributions from bend 1, drift 1, and bend 2, respectively. Due to their length, these terms are written out in Appendix~\ref{sec:formulas}.

\begin{figure}[tb!]
\centering
\includegraphics[width=0.45\textwidth]{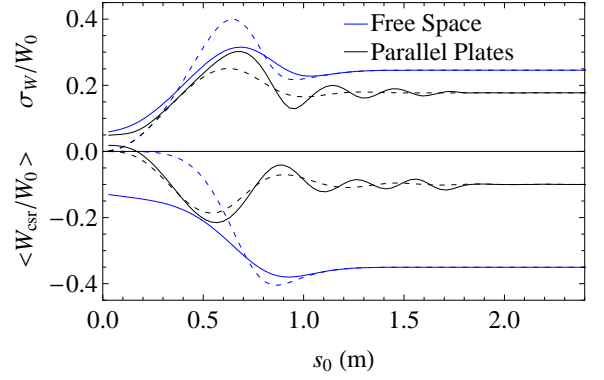}
\caption{The average energy loss and energy spread induced, per unit length, of the CSR-wake for a Gaussian bunch through the length of a bend in free space as well as between parallel plates with $H=2$cm. Solid lines have $D_1=1$m, while dashed lines have $D_1\rightarrow \infty$. Parameters used are $\kappa_1^{-1}=\kappa_2^{-1}=10$m, $\sigma_z=0.3$mm, with an energy of 1 GeV.}
\label{fg:twobendaverage}
\end{figure}

\begin{figure*}[tb!]
\centering
\includegraphics[width=\textwidth]{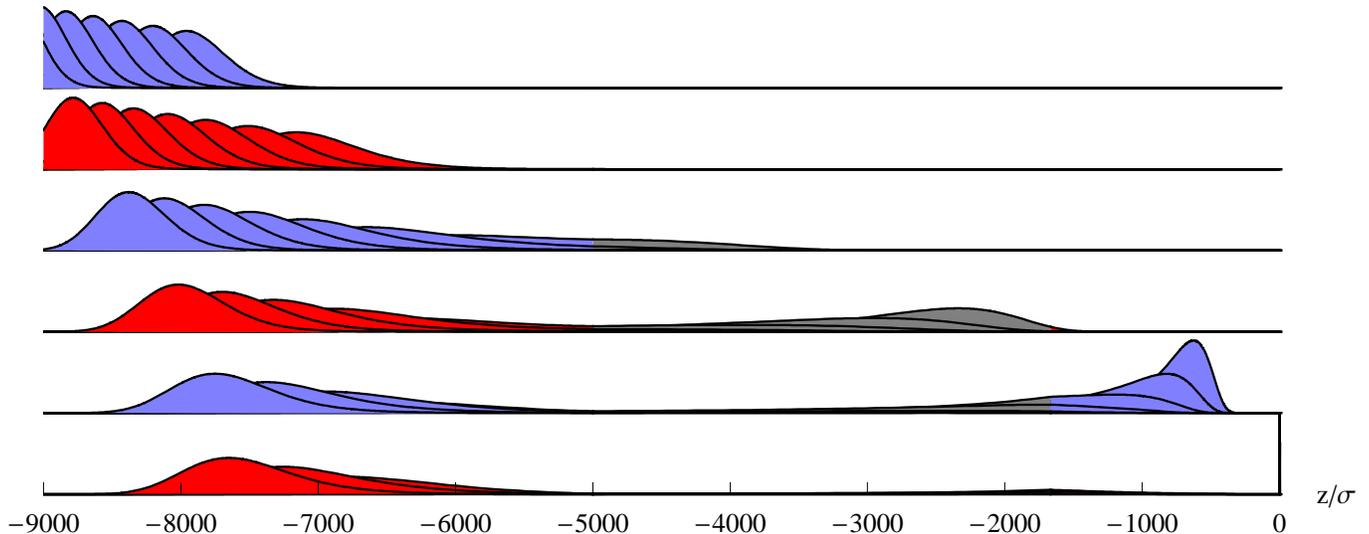}
\caption{Similar to Fig.~\ref{fg:retardedimages}, but with a 1m drift (shaded in gray) between two magnets of curvature $\kappa_1=\kappa_2=1/10$m. The center of the bunch is 50cm into the bend. A Gaussian bunch distribution is used with $\sigma_z=0.3$mm, an energy of 1GeV, and a shielding height $H=2$cm.}
\label{fg:twobendretardedimages}
\end{figure*}

A visualization of the retarded bunch and images of this geometry, similar to Fig.~\ref{fg:retardedimages}, is shown in Fig.~\ref{fg:twobendretardedimages}. Even though the bunch has progressed $50$cm into bend 1, it sees much of the retarded bunch inside bend 2, especially for test particles $z_t$ in the front of the bunch.

Two principal effects of CSR on the bunch distribution are a loss of energy and an increase in energy spread. These are calculated using the CSR-wake $W_\Csr(z)$ and the bunch distribution $\lambda(z)$, where the average energy change per unit length $\left< W_\Csr \right>$ and the standard deviation $\sigma_W\left( W_\Csr \right)$ over the distribution are
\begin{flalign}
\left<W_\Csr\right> &\equiv \int_{z_-}^{z_+} W_\Csr (z)\, \lambda(z)
\dd z, \\
\sigma_W &\equiv \left[\: \int_{z_-}^{z_+} W_\Csr^2 (z)\, \lambda(z)
\dd z- \left<W_\Csr\right>^2\right]^{1/2}.
\end{flalign}
The term $\sigma_W$ is important because it contributes to the correlated energy spread in a bunch.

To show how these quantities change as a bunch progresses through a bend, Fig.~\ref{fg:twobendaverage} plots $\left< W_\Csr \right>$ and $\sigma_W$, normalized by $W_0$, versus different bunch center coordinates $s_0$ in bend 1 using Eq.~\eqref{eq:totalbendwake} with $D_1=1$m and $\kappa_1=\kappa_2=1/10$m. In the literature, the wake near the beginning of bend 1 is often calculated as if the prior drift length $D_1\rightarrow\infty$ \cite{saldin97,agoh04}, so such calculations are plotted in dotted lines for comparison. From the difference between the two approaches, one sees the effect of bend 2, where the CSR-wake at $s_0=0$ is non-zero. In this example, they coincide after about 1.4m and 1.8m for the free space and shielded cases, respectively.

In order for it to be plausible to ignore the vacuum chamber sidewalls, such a chamber must be wide enough to allow a straight path between the retarded bunch and the test particle. In this example, the vector from a source particle at $z=-8000\,\sigma$ to the center of the bunch ($z=0$) requires that the vacuum chamber half-width must be greater than approximately 3cm.

\subsection{CSR in a Drift Between Bends}

The non-zero CSR-wake at the beginning of bend 1 in Fig.~\ref{fg:twobendaverage} is evidence that the wake in a drift region after a bend also needs to be considered. This exit-wake in the region $D_0$ following bend 1 is calculated using Eq.~\eqref{eq:1dwake} with Eq.~\eqref{eq:Xofs} and Eq.~\eqref{eq:uofs} around the center of a bunch at $s_0>B_1$. Because the bunch is moving in a straight line, the regularization procedure simply removes the need to integrate any $s'>B_1$ for the real bunch. Therefore we can use Eq.~\eqref{eq:1dwake} for bend 1, drift 1, and earlier elements, and subtract the space charge terms for $s'<B_1$. Image charges, however, still require terms for $s'>B_1$. The total exit wake is then
\begin{equation}
\begin{split}
\left.\frac{d\calE_\Csr}{ds}\right|_{\rm tot} &(s>B_1) =\left.\frac{d\calE_{\rm images}}{ds}\right|_{D_0} \\
&+ \left.\frac{d\calE_\Csr}{ds}\right|_{B_1} + \left.\frac{d\calE_\Csr}{ds}\right|_{D_1}+\cdots,
\label{eq:totalexitwake}
\end{split}
\end{equation}
where the individual terms due to element elements $D_0$, $B_1$, $D_1$, are written out in Appendix~\ref{sec:formulas}.
\begin{figure*}[t!]
\centering
\subfigure{\includegraphics[width=0.45\textwidth]{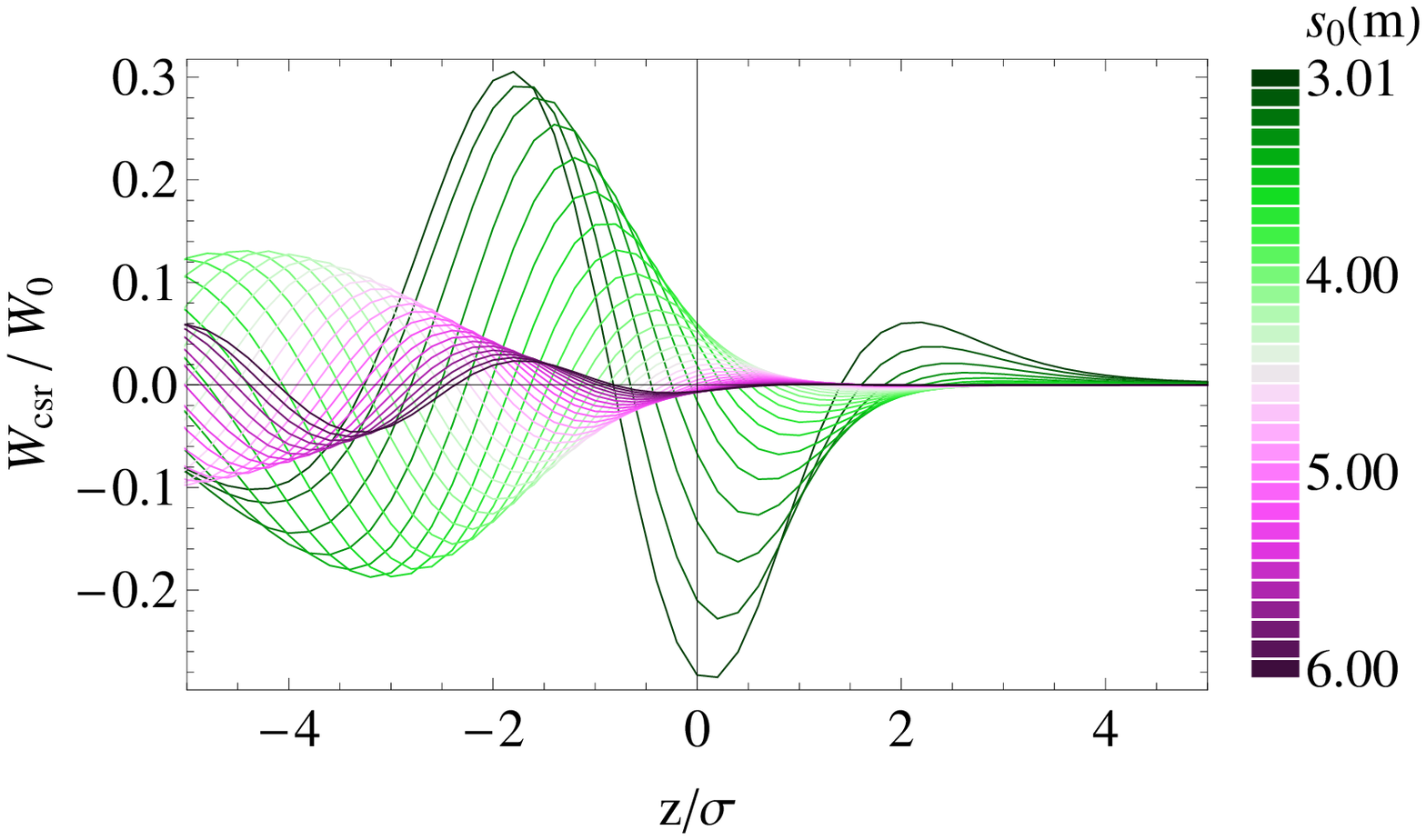}}
\subfigure{\includegraphics[width=0.45\textwidth]{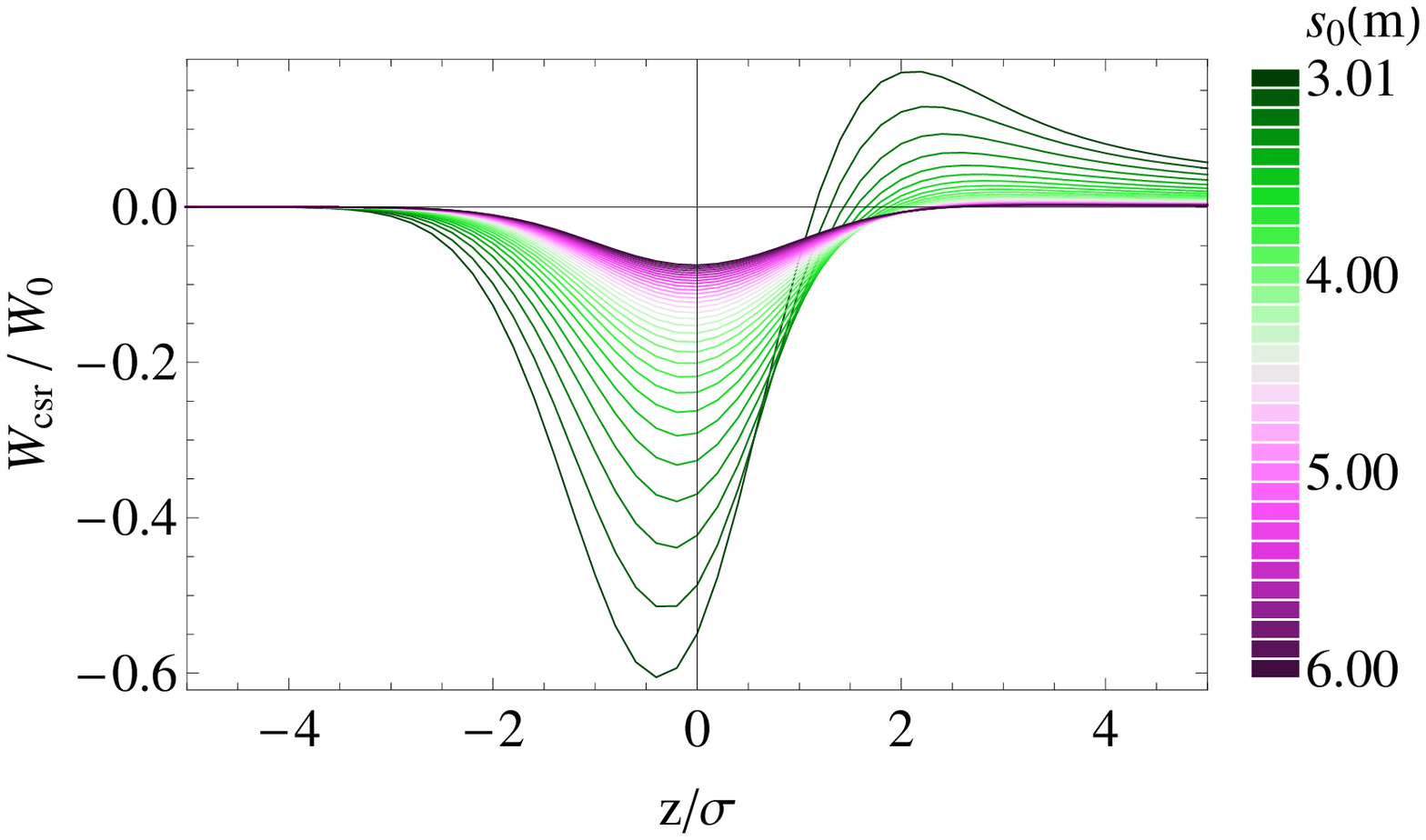}}
\caption{CSR-wakes for various bunch centers $s_0>B_1$ calculated using Eq.~\eqref{eq:totalexitwake}. The left graph uses parallel plates separated by a distance $H=2$cm, while the right graph is for free space ($n=0$ terms only in Eqs.~\eqref{eq:exitB1}-\eqref{eq:exitD1}, and without Eq.~\eqref{eq:exitD0}). The bending radius $\kappa_1^{-1}=10$m, and the bunch has a Gaussian profile with $\sigma=0.3$mm and an energy of 1GeV.}
\label{fg:exitwakes}
\end{figure*}
\begin{figure*}[h!]
\centering
\subfigure{\includegraphics[width=0.45\textwidth]{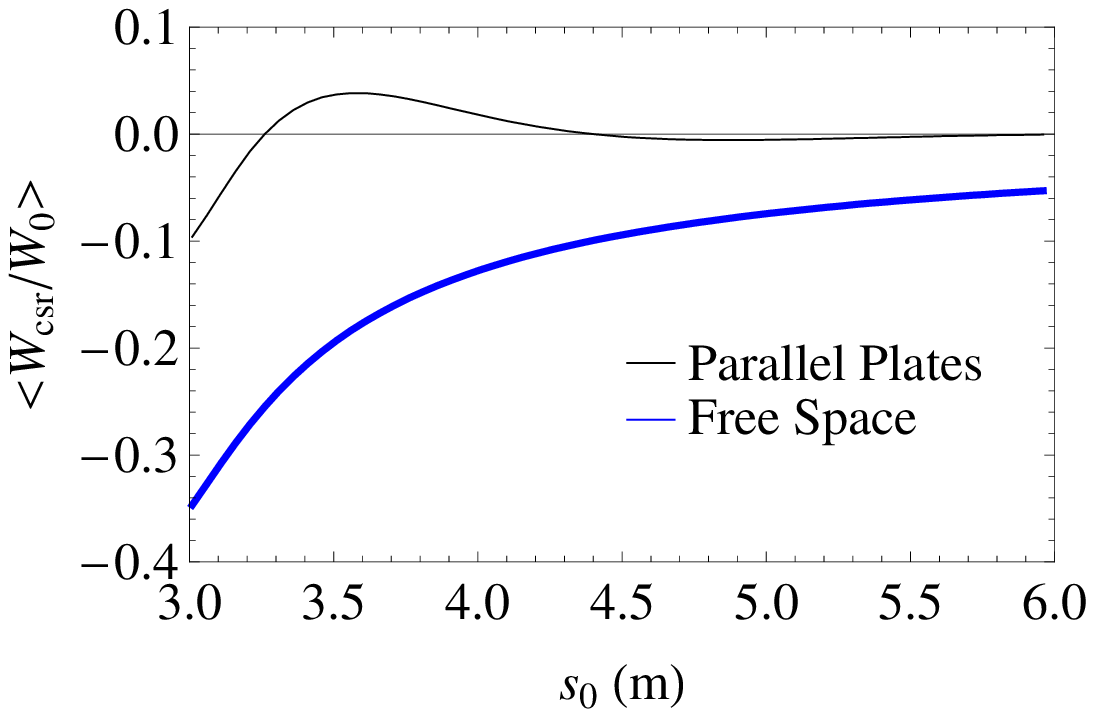}}
\subfigure{\includegraphics[width=0.45\textwidth]{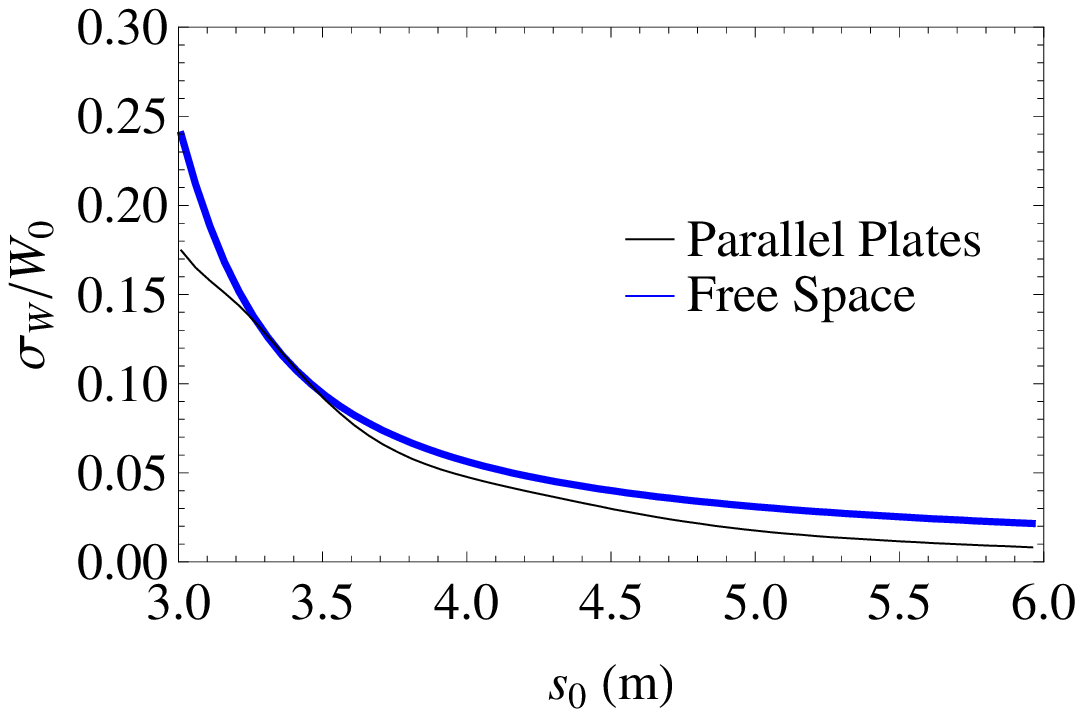}}
\caption{The average energy loss and energy spread per unit length of the exit wakes in Fig.~\ref{fg:exitwakes}. In this example, shielding by parallel plates drastically reduces the energy loss, but only marginally reduces the energy spread, when compared to free space calculations.}
\label{fg:exitaveragerms}
\end{figure*}
For a magnet of length $B_1 =3$m, the exit wakes in the following drift $D_0$ are shown in Fig.~\ref{fg:exitwakes} for bunch centers in the following 3 meters between parallel plates and in free space. The average and standard deviation of the wakes through this region are shown in Fig.~\ref{fg:exitaveragerms}. In the shielded situation, one sees that the bunch actually gains some energy in a short length following the bend, and that the total energy loss between parallel plates is negligible compared to the free space losses. Energy spread, however, is qualitatively the same in both cases.



\section{Bunch Compression}

Bunch compression or decompression can be achieved in a bending magnet if there is a correlation between energy and longitudinal position of particles in the bunch. To exactly calculate CSR for this, however, requires at least a 2-dimensional model, because particles of different energies travel on different orbits. In the framework of the 1-dimensional model described by Eq.~\eqref{eq:1dwake}, this effect can be approximately modeled by allowing the bunch length to be time dependent, and neglecting variations in the velocity $\beta\,c$. The density and current are then
\begin{equation}
\begin{split}
\rho(s, t) &= Q\,  \frac{1}{\sigma(t)} \widetilde{\lambda} \left( \frac{ s -s_b  -\beta \,c t}{\sigma(t)} \right), \\
\BfJ(s,t)  &=  Q\,  \beta\, c \,  \frac{1}{\sigma(t)} \widetilde{\lambda} \left( \frac{ s -s_b  -\beta \,c t}{\sigma(t)} \right),
\end{split}
\label{eq:rhocompress}
\end{equation}
where $\widetilde{\lambda}$ has unit norm and variance with respect to $s$, as in Eq~\eqref{eq:lambdatilde}. The time derivative of $\rho(s,t)$ is
\begin{equation}
\begin{split}
\frac{\partial}{\partial t} \rho(s,t) =& -\beta\,  c\, \frac{Q\, \widetilde{\lambda'} \left( \dfrac{s_t }{\sigma} \right)}{\sigma^2} \\
&-\dot{\sigma}\left[\frac{Q\,\widetilde{\lambda} \left( \dfrac{ s_t}{\sigma} \right)}{\sigma^2}+\frac{s_t }{\sigma}\frac{Q\,\widetilde{\lambda'} \left( \dfrac{ s_t}{\sigma} \right)}{\sigma^2} \right].
\end{split}
\end{equation}
with $s_t \equiv s -s_b  -\beta \,c t$. Note that $\dot{\sigma} /(\beta\,c)$ is on the order of $\sigma/B$ in a magnet of length $B$, and $(s -s_b  -\beta \,c t)$ is on the order of $\sigma$ for all relevant $(s,t)$, and therefore the term in brackets is on the order of $\sigma/B \ll 1 $ relative to the first term, and will be neglected. With such an approximation, the CSR-wake in a bunch compression system can be modeled by simply making the substitutions
\begin{align}
\lambda(s_r)&\rightarrow \frac{1}{\sigma(t_{\rm ret})}\widetilde{\lambda} \left(\frac{s_r}{\sigma(t_{\rm ret})}\right) \label{eq:lambdacompress}\\
\lambda'(s_r)&\rightarrow \frac{1}{\left[\sigma(t_{\rm ret})\right]^2} \, \widetilde{\lambda}' \left(\frac{s_r}{\sigma(t_{\rm ret})}\right)
\label{eq:lambdaprimecompress} \\
t_{\rm ret} &= t-\sqrt{r^2+(n \, H)^2}/c)
\end{align}
in all of the previous formulas, with $r = \| \BfX(s)-\BfX(s')\|$, as in Eq.~\eqref{eq:1dwake}. This accounts for the real charges ($n=0$) and image charges ($n\neq 0$) at the appropriate retarded times.

\begin{figure*}[tb!]
\centering
\subfigure[Free Space Bunch Compression]{\label{fg:compressionaveragermsA}\includegraphics[width=0.45\textwidth]{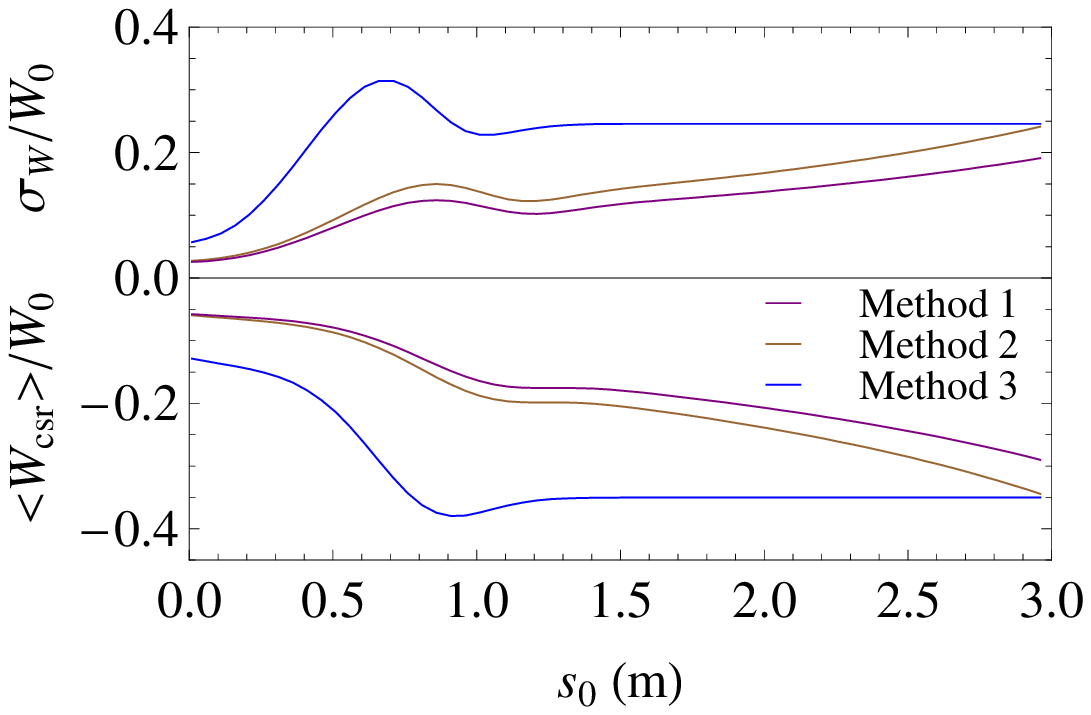}}
\subfigure[Parallel Plates Bunch Compression]{\label{fg:compressionaveragermsB}\includegraphics[width=0.45\textwidth]{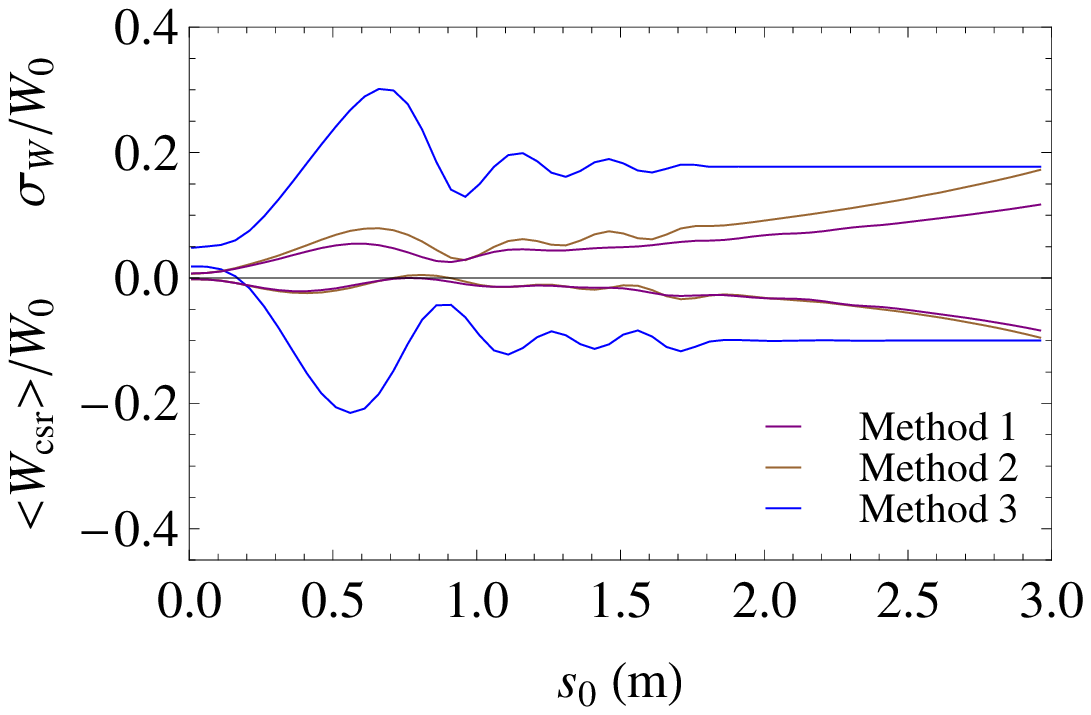}}\\
\subfigure[Free Space Bunch Compression, $D_1\rightarrow\infty$ ]{\label{fg:compressionaveragermsinfinityA}\includegraphics[width=0.45\textwidth]{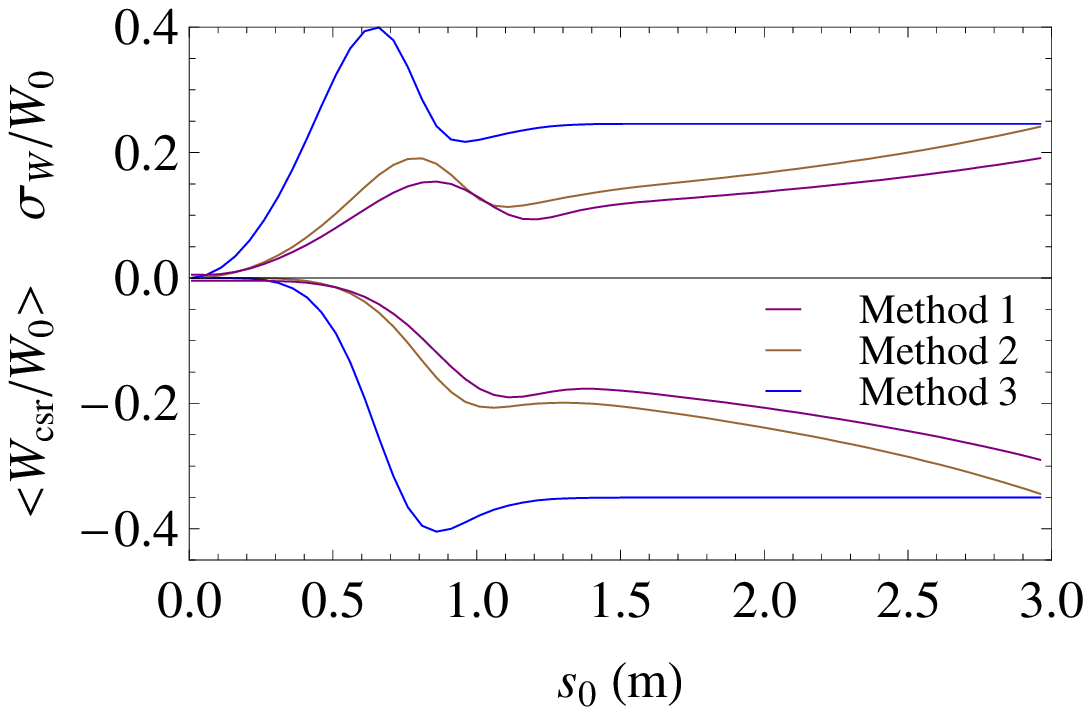}}
\subfigure[Parallel Plates Bunch Compression, $D_1\rightarrow\infty$]{\label{fg:compressionaveragermsinfinityB}\includegraphics[width=0.45\textwidth]{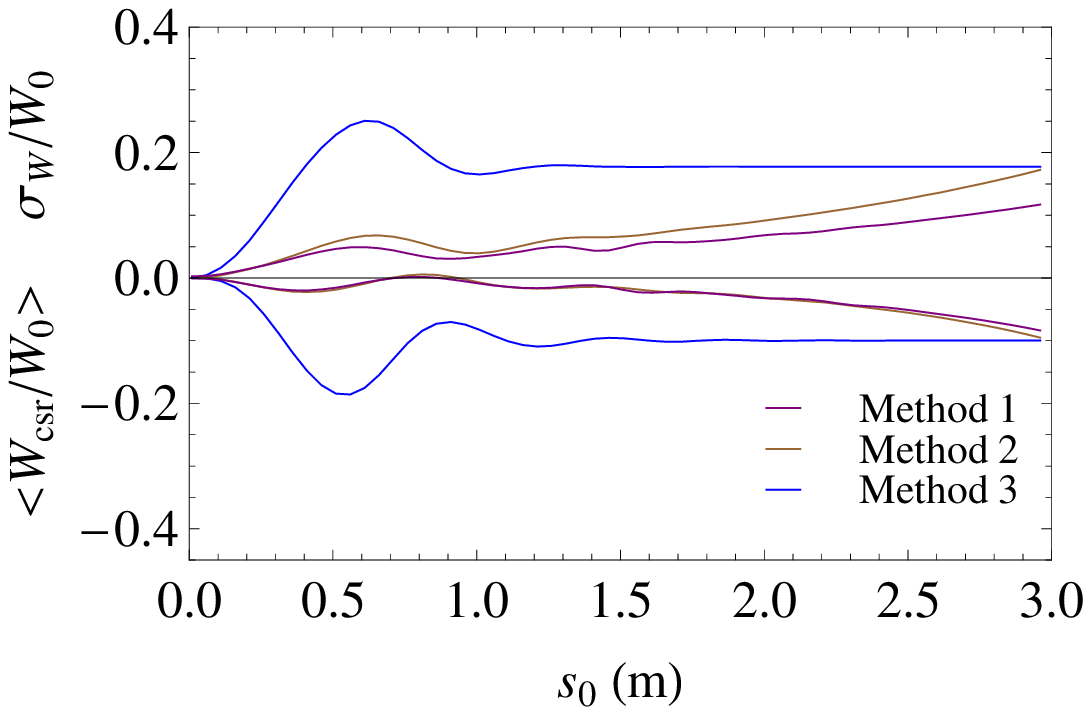}}
\caption{The average and standard deviation  of the CSR-wake in free space (Fig.~\ref{fg:compressionaveragermsA}) and between parallel plates with $H=2$cm (Fig.~\ref{fg:compressionaveragermsB}) over a Gaussian bunch, compressing from $\sigma=0.9$mm to $\sigma=0.6$mm linearly through bend 2, and from $\sigma=0.6$mm to $\sigma=0.3$mm linearly through bend 1, using methods described in the text. Figures~\ref{fg:compressionaveragermsA}--\ref{fg:compressionaveragermsB} have $D_1=1$m, while Figs.~\ref{fg:compressionaveragermsinfinityA}--\ref{fg:compressionaveragermsinfinityB} have $D_1\rightarrow\infty$. The lengths $B_1=B_2=3$m, the bending radii are $\kappa_1^{-1}=\kappa_2^{-1}=10$m, and the energy is 1GeV. }
\label{fg:compressionaveragerms}
\end{figure*}

Calculations for the average and standard deviation of the CSR-wake with a linearly compressing bunch through the length of bend 1 in free space are shown in Fig.~\ref{fg:compressionaveragermsA}. The above approximation is referred to as Method 1. Method 2 calculates the instantaneous CSR-wake  of a compressing bunch at each point in the bend \emph{as if} it always had its instantaneous length. Such a scheme is essentially what particle tracking codes (e.g.\ {\tt elegant}~\cite{borland01}, Bmad~\cite{b:bmad}) use for CSR simulation. For reference, Method 3 calculates the CSR-wake for a non-compressing bunch that maintains the same length as the final compressed length in Methods 1 and 2. In this example, Method 2 overestimates the CSR effect compared to the more realistic Method 1, and both exhibit a much smaller effect than Method 3. At the end of the magnet ($s_0 = 3$m), the CSR-wake, according to Method 1, has yet to reach its corresponding steady-state strength.

Figure~\ref{fg:compressionaveragermsB} shows these same calculations but between parallel plates with $H=2$cm. One sees that the energy loss in method 2 is similar to that in method 1, but the energy spread induced is overestimated. Free space and shielded calculations are repeated with $D_1\rightarrow\infty$ in Figs.~\ref{fg:compressionaveragermsinfinityA}--\ref{fg:compressionaveragermsinfinityB}, which when compared with Figs.~\ref{fg:compressionaveragermsA}--\ref{fg:compressionaveragermsB} one can see the effect of the previous bend $B_2$.

\section{Coherent Power Spectrum}
Some of the first CSR calculations are found in an originally unpublished report by Schwinger \cite{schwinger45}. Here we use one of his methods to derive an exact expression for the coherent energy loss by a Gaussian beam, which is then used to verify our earlier calculations.  Consider the power spectrum due to a \emph{single} particle moving on a circle with velocity $\beta\,c$, which is proportional to the absolute square of the Fourier transform electric field $\BfE^{(1)}(\Omega, t)$, integrated over solid angle $\Omega$, as in
\begin{equation}
\frac{d P^{(1)}}{d \omega} \propto \int \dd \Omega \left|\int_{-\infty}^{\infty}\dd t \:e^{i \omega t} \BfE^{(1)}(\Omega,t) \right|^2.
\end{equation}
For $N$ particles moving on this circle with positions $s=s_n+\beta\,ct$, the total electric field can be written in terms of the single particle's electric field ($s_n=0$), as in
\begin{equation}
\BfE^{(N)}(\Omega, t) = \sum_{n=1}^N \BfE^{(1)}(\Omega, t-t_n),
\end{equation}
where the time deviations $t_n=s_n/(\beta\, c)$. By changing variables, this means that the $N$ particle power spectrum is simply
\begin{equation}
  \frac{d P^{(N)}}{d\omega} = \left|
  \sum_{n=1}^{N} e^{i \omega t_n} \right|^2 \frac{d P^{(1)}}{d\omega}.
  \label{eq:discretePN}
\end{equation}
These phase factors can be separated into terms with $m=n$ and $m\neq n$,
\begin{equation}
\begin{split}
&\frac{d P^{(N)}}{d\omega}   = \left(\sum_{m=1}^{N} e^{i \omega t_m } \sum_{n=1}^{N} e^{-i \omega t_n } \right)
  \frac{d P^{(1)}}{d\omega} \\
  &=  N\frac{d P^{(1)}}{d\omega} \\
  &\quad+ \frac{d P^{(1)}}{d\omega} \sum_{m=1}^N  \Exp{i\, \omega \frac{ s_m}{\beta c}}\sum_{\substack{n=1\\n \neq m}}^N \Exp{-i\, \omega \frac{ s_n}{\beta c}}
 ,
\end{split}
\end{equation}
so that the second term can be written as a correlation between different particles
\begin{equation}
\begin{split}
\sum_{m \neq n}&\Exp{i\, \omega \frac{ s_m - s_n}{\beta\,c}} \simeq N(N-1) \\
& \times \int \!\!\! \dd s\: \lambda(s)\Exp{i\,\omega \frac{s}{\beta c}} \\
& \times \int \!\!\! \dd s'\: \lambda(s') \Exp{-i\,\omega \frac{s'}{\beta c}}
\end{split}\end{equation}
using the normalized particle distribution $\lambda(s)$ along the circle. The $N$ particle power spectrum is then
\begin{equation}
\begin{split}
&\frac{d P^{(N)}}{d\omega}(\omega) \simeq  \underbrace{N  \frac{d P^{(1)}}{d\omega}}_{\rm incoherent}\\
 &\quad+\underbrace{N (N-1)\left| \int \!\!\! \dd s \, \lambda (s) \exp
  \left(i\, \frac{\omega s}{\beta c}\right) \right|^2 \frac{d P^{(1)}}{d\omega}}_{\rm coherent}.
 \label{eq:PNspectrum}
\end{split}\end{equation}
The first term in Eq.~\eqref{eq:PNspectrum} is the incoherent power spectrum, while
the second is the coherent power spectrum. The squared integral is called the form-factor.


\begin{figure}[t!]
\centering
\includegraphics[width=0.45\textwidth]{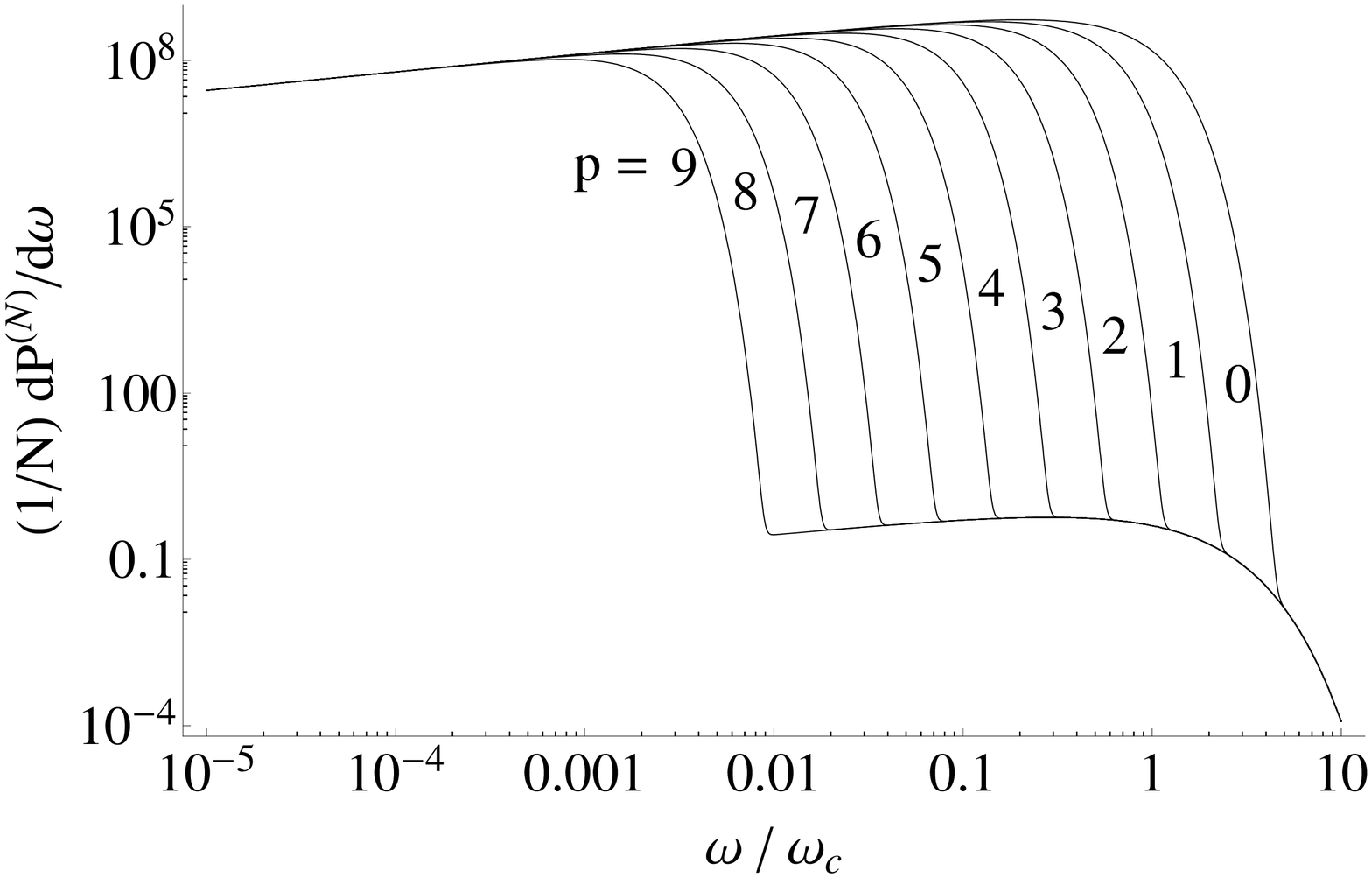}
\caption{The power spectrum in Eq.~\eqref{eq:PNspectrum}, per particle, using a Gaussian form factor with various values of the coherence parameter $a_c=2^p$, defined in Eq.~\eqref{eq:ac}. The lower frequencies are enhanced by a factor of $N$, and in this example $N=10^9$.}
\label{fg:gaussianspectrum}
\end{figure}

In free space, the well-known single particle power spectrum is
\begin{equation}
\frac{d P^{(1)}}{d\omega}(\omega) = \frac{P^{(1)}}{\omega_c} \, S\left( \frac{\omega}{\omega_c} \right),
\label{eq:P1spectrum}
\end{equation}
where $\omega_c \equiv \frac{3}{2} \gamma^3\, c\, \kappa$ is the critical frequency \cite{jackson99,handbook}. The function $S$ is defined as
\begin{equation}
S(\xi) \equiv \frac{9\sqrt{3}}{8\pi}\,\xi\int_{\xi}^\infty\dd x\: K_{5/3} (x),
\end{equation}
in which $K$ is a modified Bessel function. The integral $\int_0^\infty S(x)\dd x =1$, and the total power lost by a single particle is
\begin{equation}
  P^{(1)} \equiv \frac{2}{3} r_c\, m\, c^3\, \beta^4\, \gamma^4\, \kappa^2.
\end{equation}
For a Gaussian distribution with variance $\sigma^2$ and  $\kappa\, \sigma \ll 1$, the form-factor is, extending the integration limits to infinity,
\begin{equation}
\begin{split}
\left| \int_{-\infty}^{\infty} \!\!\!\dd s \, \frac{\Exp{i\, \dfrac{\omega s}{\beta c}-\dfrac{s^2}{2\sigma^2}}}{\sqrt{2 \pi}\, \sigma}  \right|^2 &= \Exp{-\frac{\sigma^2\,\omega^2}{\beta^2\,c^2}}\\
=& \exp \left\{-\left(a_c\frac{\omega}{\omega_c}\right)^2\right\},
\end{split}
\end{equation}
defining the coherence factor
\begin{equation}
\begin{split}
a_c &\equiv \frac{3}{2\beta}\gamma^3\, \kappa\, \sigma
\label{eq:ac}\\
 &= \frac{\sigma}{\beta\,c}\omega_c.\
\end{split}
\end{equation}
The total power spectrum per particle for an $N$-particle Gaussian distribution with various values of $a_c$ is shown in Fig.~\ref{fg:gaussianspectrum}. One sees from the exponential that the lower frequencies, up to a cutoff frequency around $\omega = \beta\,c/\sigma$, are enhanced by a factor of $N$ by the coherent part of Eq.~\eqref{eq:PNspectrum}. The spectrum at higher frequencies agrees with the familiar single particle spectrum in Eq.~\eqref{eq:P1spectrum}.

It turns out that Eq.~\eqref{eq:PNspectrum} can be integrated exactly for a Gaussian distribution. Explicitly, the total power radiated by $N$ particles is
\begin{equation}
\begin{split}
&P^{(N)} = N P^{(1)} \int_0^\infty S(x)\dd x + N(N-1)P^{(1)}\\
 &\qquad\qquad\quad\times\frac{9\sqrt{3}}{8\pi}\int_0^\infty x\, e^{-a_c^2\, x^2}\left[\int_x^\infty\!\! K_{5/3} (y)\dd y \right]\!\! \dd x \\
&= N P^{(1)} +N(N-1) P^{(1)}\\
  &\qquad\qquad\quad\times\frac{9\sqrt{3}}{8\pi}\int_0^\infty K_{5/3}(y) \left[ \int_0^y \!\! x\, e^{-a_c^2\, x^2} \dd x\right]\!\!\dd y\\
&= N P^{(1)}  + N(N-1)\,P^{(1)}\, T_c\left(\frac{3}{2\beta}\gamma^3\, \kappa\, \sigma \right),
\end{split}
\label{eq:PNGaussian}
\end{equation}
in which the final integral yields the coherence function defined as
\begin{equation}
  T_c(a_c) \equiv  \frac{9}{32 \sqrt{\pi}\, a_c^3}\Exp{
  \frac{1}{8 a_c^2}} K_{5/6}\left(\frac{1}{8 a_c^2}\right)-\frac{9}{16  a_c^2}.
  \label{eq:Tc}
\end{equation}

\begin{figure}[tb!]
\centering
\includegraphics[width=0.45\textwidth]{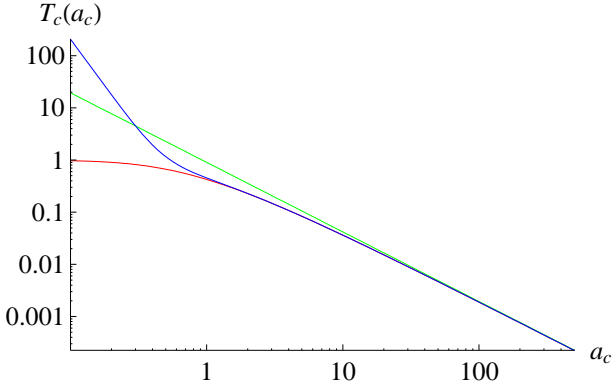}
\caption{The coherence function $T_c(a_c)$ of Eq.~\eqref{eq:Tc} is plotted in red. The green curve is the first term in the asymptotic expansion in Eq.~\eqref{eq:Tas} \cite{nodvicksaxon54}, and the blue curve uses all three terms in Eq.~\eqref{eq:Tas}. }
\label{fg:Tcompare}
\end{figure}

The limit $\lim_{a\to 0^+} T_c(a) = 1$, which is to say that an infinitely narrow bunch radiates as one charge. In practical situations $a_c\gg 1$, so an asymptotic expansion of $T_c$ gives the useful approximation
\begin{equation}
\begin{split}
T_c(a_c) \sim& \frac{9\, \Gamma \left( \frac{5}{6}\right)}{16^{2/3}\sqrt{\pi}} \left(\frac{1}{a_c}\right)^{4/3} -\frac{9}{16}\left(\frac{1}{a_c}\right)^{2}\\
 &+\frac{9\, \Gamma \left( \frac{5}{6}\right)}{32\cdot 2^{2/3}\sqrt{\pi}} \left(\frac{1}{a_c}\right)^{10/3} + \ldots \ \ .
\label{eq:Tas}
\end{split}\end{equation}

The first term in Eq.~\eqref{eq:Tas} is given in Ref.~\cite{nodvicksaxon54}. Figure~\ref{fg:Tcompare} compares this first term to the exact expression in Eq.~\eqref{eq:Tc} and to all three terms in Eq.~\eqref{eq:Tas}. One sees an excellent approximation for $a_c\gtrsim 50$ using the first term and for $a_c\gtrsim 1$ using all three terms in Eq.~\eqref{eq:Tas}. Also, the average coherent energy lost per particle per unit length is
\begin{flalign}
\left<\frac{P^{(N)}}{N\,\beta c}\right>_{\rm coh.}\!\!\!\! &=\! -\frac{2}{3}(N-1) r_c m c^2  \gamma^4 \beta^3 \kappa^2T_c\left(\frac{2}{3}\gamma^3\kappa \sigma\! \!\right) \label{eq:avEexact} \\
&\sim -\frac{\Gamma\left(\frac{5}{6}\right)}{6^{1/3}\sqrt{\pi}} W_0 + \ldots\ \ , \label{eq:avEas}
\end{flalign}
using $W_0$ defined in Eq.~\eqref{eq:W0}. The numerical coefficient $\Gamma(5/6)\, 6^{-1/3}\, \pi^{-1/2}\simeq 0.350$. The same procedure in Eq.~\eqref{eq:PNGaussian} and  Eq.~\eqref{eq:Tas} can be carried out for a uniform distribution of length $\Delta L$ with the same variance $\sigma^2$, implying that $\Delta L = 2\sqrt{3}\,\sigma$. The result yields the same form as Eq.~\eqref{eq:avEas}, except with the numerical coefficient $2^{-4/3}\simeq 0.397$. This term was originally derived in Ref.~\cite{schwinger45}.

\begin{figure}[tb!]
\centering
\includegraphics[width=0.45\textwidth]{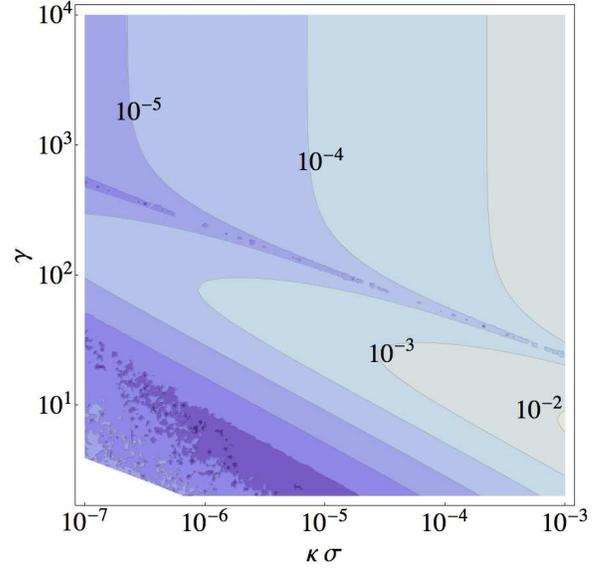}
\caption{The relative difference $|(b-a)|/b$ for $a$ the average energy lost using Eq.~\eqref{eq:Esteadystate}, and $b$ being the result in  Eq.~\eqref{eq:avEexact}.}
\label{fg:energyerror}
\end{figure}

To verify that the CSR-wake does indeed represent the coherent energy lost, the relative difference of the average energy loss using the steady-state wake of a Gaussian bunch in Eq.~\eqref{eq:Esteadystate} to the result in Eq.~\eqref{eq:avEexact} is plotted in Fig.~\ref{fg:energyerror}. One sees that the relative difference is at most $1\%$ in this practical parameter range, and that occurs with relatively long bunches. We speculate that this error is caused by the regularization procedure that subtracts the space charge term from the longitudinal electric field.

\begin{figure}[tb!]
\centering
\includegraphics[width=0.45\textwidth]{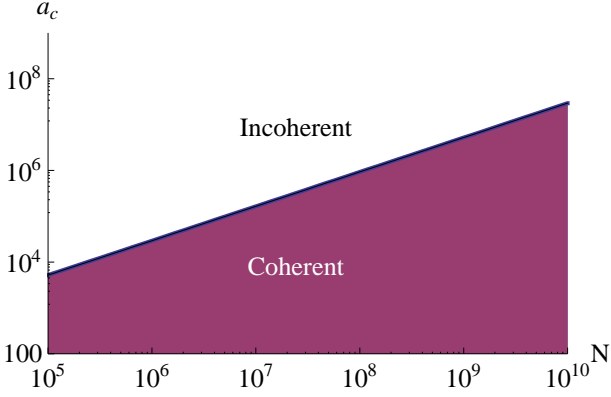}
\caption{The dividing line where the coherent power equals the incoherent power, \emph{i.e.}\ the total power is twice the incoherent power. Below this line, the coherent power dominates the total power.}
\label{fg:coherence}
\end{figure}

The relevance of the coherence function depends on the number of particles $N-1\simeq N$. The coherent power radiated equals the incoherent power radiated when $N\cdot T_c\,(a_c) = 1$, illustrated in Fig.~\ref{fg:coherence}. Using Eq.~\eqref{eq:Tas}, the coherent power dominates the total power when
\begin{equation}
\kappa\,\sigma\: \lesssim\: \frac{N^{3/4}}{\gamma^3}.
\end{equation}

\section{Conclusion}
The wake-field due to CSR of a 1-dimensional bunch traveling on a curve without small angle or high energy approximations has been derived using Jefimenko's forms of Maxwell's equations. This exact solution allowed us to quantify the accuracy of the approximations of the steady-state CSR-wake in a bend given in Ref.~\cite{saldin97} and Ref.~\cite{sagan08} showing that the former is inaccurate at low energies and long bunch lengths, and that the latter is much more accurate down to low energies. All approximations tend to overestimate the CSR-wake. For planar orbits the equations are extended to include shielding by perfectly conducting parallel plates using the image charge method.

The formulas have been applied to the geometry of a bend preceded by a drift, preceded by another bend, and show that the CSR-wake well inside the downstream bend is influenced by the upstream bend for the parameters used. In fact, a bunch near the entrance of a bend is influenced by the CSR-wake due to the previous bend much more than by that due to the previous drift. Shielding by parallel plates reduces the energy loss rate significantly, but the effect on reducing energy spread increase is far less dramatic, in both the drift and bend regions.

Bunch compression has been added to this model by allowing the bunch length to be time dependent, so that the retarded charge density seen by a test particle is appropriately taken into account. This method has been compared to simple methods used by particle simulation codes Bmad and {\tt elegant}, and it is shown that these tend to overestimate the effect \cite{b:bmad,borland01}.

Additionally, an exact expression for the coherent power lost by a 1-dimensional Gaussian bunch moving in a circle has been derived by integrating the power spectrum, following the method of Schwinger \cite{schwinger45}. When compared to the energy loss rate by the CSR-wake, the two show slight deviations. This could be due to the regularization procedure for the 1-dimensional CSR-wake that subtracts off the space charge term.

\appendix
\section{Formulas for Multiple Bends and Drifts}
\label{sec:formulas}

For ease of reading, the individual terms terms in Eq.~\eqref{eq:totalbendwake} and Eq.~\eqref{eq:totalexitwake} have been deferred to here. They are calculated by applying Eq.~\eqref{eq:1dwake}, regularized by Eq.~\eqref{eq:ESC}, including image charges as in Eq.~\eqref{eq:imagewake}, to the geometry in Eq.~\eqref{eq:Xofs}.

In Eq.~\eqref{eq:totalbendwake}, the first term $d \calE_{\rm csr} /ds|_{B_1}$ is the sum of Eq.~\eqref{eq:ECSRbend} and Eq.~\eqref{eq:intbend2image} with $\kappa \rightarrow \kappa_1$ and $\theta \rightarrow \kappa_1\,s$, explicitly
\begin{widetext}
\begin{equation}
\begin{split}
\left.\frac{d\calE_{\Csr}}{ds}(s)\right|_{B_1}  =&N r_c m c^2  \left \{ \int_{\alpha_a}^{\alpha_b} \dd \alpha \:   \left(\frac{\beta^2 \Cos{ \alpha}-1}{2|\Sin{\alpha/2}|}+ \frac{1}{\gamma^2}\frac{\sgn(\alpha)-\beta\,\Cos{\alpha/2}}{ \alpha-2\beta|\Sin{\alpha/2}|} \right)\lambda' (s_\alpha) \right.  \\
&\left.\left. -\frac{\kappa_1 \, \lambda(s_\alpha)}{2|\Sin{\alpha/2}|}\right|_{\alpha_a}^{\alpha_b}+ \int_{\Delta_{a}}^{\infty} \dd \Delta \: \frac{1}{\gamma^2}\frac{\lambda'(z-\Delta)}{\Delta} + \int_{\Delta_b}^{\infty} \dd \Delta \: \frac{1}{\gamma^2}\frac{\lambda'(z+\Delta)}{\Delta}  \right. \\
&\left.+\sum_{n=1}^{\infty} 2 (-1)^n \left[ \left. \frac{-\kappa_1 \, \lambda(s_{\alpha,n})}{r_{\alpha,n}}\right|_{\alpha_a}^{\alpha_b} +\int_{\alpha_a}^{\alpha_b} \dd \alpha \:\frac{\beta^2 \Cos{ \alpha}-1}{r_{\alpha,n}}\lambda' (s_{\alpha, n}) \right]\right \}
\end{split}
\end{equation}
with the definitions
\begin{equation}
\begin{split}
\alpha_a &\equiv \kappa_1(s-B_1),\\
\alpha_b &\equiv \kappa_1 \, s,\\
\Delta_{a} &\equiv s - 2\beta \frac{1}{\kappa_1}\Sin{\frac{\kappa_1\, s}{2}},\\
\Delta_{b} &\equiv B_1-s +2\beta \frac{1}{\kappa_1}\Sin{\frac{\kappa_1( B_1 -s)}{2}},\\
r_{\alpha,n}&\equiv\sqrt{2-2\cos \alpha + (n \,\kappa_1 H)^2},\\
s_\alpha &\equiv s-s_0- \frac{1}{\kappa_1}\left(\alpha - \beta \sqrt{2-2\cos \alpha} \right),\\
s_{\alpha,n} &\equiv  s-s_0-\frac{1}{\kappa_1}\left(\alpha - \beta\, r_{\alpha,n} \right).
\end{split}
\end{equation}
Some trigonometric functions have been simplified, and the space charge integrals have changed variables to $\Delta = (\alpha - \beta\sqrt{2-2\cos\alpha})/\kappa_1$. These terms account for the regularized CSR-wake and image charges in bend 1. The next terms are
\begin{equation}
\begin{split}
\left.\frac{d\calE_{\Csr}}{ds}\right|_{D_1} &=N r_c m c^2 \int_{0}^{D_1}  \dd   L \sum_{n=0}^{\infty} (2-\delta_{n,0}) (-1)^n \left \{ \frac{T_L}{R_{L,n}^3}\, \lambda(s_{L,n})+ \left[ \beta^2\frac{\Cos{\kappa_1 \, s}}{R_{L,n} } -\beta \frac{T_L }{R_{L,n}^2} \right]\lambda'(s_{L,n})  \right \}  \\
R_{L,n} &\equiv \frac{1}{\kappa_1}\sqrt{2-2\Cos{\kappa_1 \, s}+2 \kappa_1 L \Sin{\kappa_1 \, s} + (\kappa_1 L)^2 +(\kappa_1 n H)^2} \\
T_L &\equiv L \Cos{\kappa_1 \, s} +\frac{1}{\kappa_1}\Sin{\kappa_1 \, s} \\
s_{L,n} &\equiv  -L -s_0 + \beta R_{L,n}
\end{split}
\end{equation}
\begin{equation}
\begin{split}
\left.\frac{d\calE_{\Csr}}{ds}\right|_{B_2} &=N r_c m c^2 \int_{0}^{B_2}  \dd   L \sum_{n=0}^{\infty} (2-\delta_{n,0}) (-1)^n \left \{ \frac{T_L}{R_{L,n}^3}\, \lambda(s_{L,n}) +\left[ \beta^2\frac{\Cos{\kappa_1 \, s+ \kappa_2\, L}}{R_{L,n}} -\beta \frac{T_L }{R_{L,n}^2} \right]\lambda'(s_{L,n})  \right \}  \\
R_{L,n} &\equiv \sqrt{\left(\frac{\Cos{\kappa_1 \, s} -1}{\kappa_1}+\frac{1-\Cos{ \kappa_2 \, L}}{\kappa_2}\right)^2+\left(D_1+\frac{\Sin{\kappa_1 \, s}}{\kappa_1}+\frac{\Sin{ \kappa_2\, L}}{\kappa_2}\right)^2 +\left(n\,H\right)^2}\\
T_L &\equiv D_1 \Cos{\kappa_1 \, s} +\frac{\kappa_2-\kappa_1}{\kappa_1\kappa_2}\Sin{\kappa_1 \, s} +\frac{1}{ \kappa_2}\Sin{\kappa_1 \, s+ \kappa_2\, L}\\
s_{L,n}&\equiv -L -D_1 -s_0  + \beta\, R_{L,n} .
\end{split}
\end{equation}
Note that the lower limit of the sums have been set to $n=0$ to account for the real charges as well as image charges, necessitating the use of Kronecker's delta.  Alternatively, if only free space terms are desired, the above formulas can be used with the $n=0$ term only. The dummy variable $s'$ has been rescaled to $L$ which integrates backwards over the length of the appropriate element. The terms $R_{L,n}$, $T_L$, and $s_{L,n}$ are redefined after each equation in order to keep the naming sane.

Similarly, the wake at $s>B_1$ after bend, as in Eq.~\eqref{eq:totalexitwake}, contains the terms
\begin{equation}
\begin{split}
\left.\frac{d\calE_\Csr}{ds}\right|_{D_0}= -N r_c m c^2 &\left\{ \int_{-\infty}^{B_1} \dd s'\left[ \frac{1}{(s-s')^2}\lambda(s'-s_0+\beta(s-s')) +\beta\frac{\beta-1}{s-s'}\lambda'(s'-s_0+\beta(s-s'))\right] \right.\\
 &+ \sum_{n=1}^{\infty} 2(-1)^n \frac{\lambda(B_1-s_0+\beta\sqrt{(s-B_1)^2+(n H)^2})}{\sqrt{(s-B_1)^2+(n H)^2}} \\
& \left. +\sum_{n=1}^{\infty} 2(-1)^n\int_{0}^{\infty}  \dd   L \frac{\lambda'(L+B_1-s_0+\beta\sqrt{(s-B_1-L)^2+(n H)^2})  }{\gamma^2 \sqrt{(s-B_1-L)^2+(n H)^2}}  \right\},
\end{split}
\label{eq:exitD0}
\end{equation}
\begin{equation}
\begin{split}
\left.\frac{d\calE_\Csr}{ds}\right|_{B_1} &= N r_c m c^2\sum_{n=0}^{\infty} (2-\delta_{n,0})(-1)^n  \int_0^{B_1} \dd L\left \{\frac{T_L}{R_{L,n}^3}\lambda(s_{L,n})+\left[\beta^2 \frac{\Cos{\kappa_1 L}}{R_{L,n}}-\beta \frac{T_L}{R_{L,n}^2} \right]\lambda'(s_{L,n})\right\}\\
R_{L,n} &\equiv \sqrt{\frac{2-2\Cos{\kappa_1 L}}{\kappa_1^2}+2\frac{(s-B_1)\Sin{\kappa_1 L }}{\kappa_1}+(s-B_1)^2+\left(n\,H\right)^2} \\
T_L &\equiv s-B_1 +\frac{1}{\kappa_1}\Sin{\kappa_1 L} \\
s_{L,n} &\equiv -L+B_1-s_0 + \beta \,R_{L,n},
\end{split}
\label{eq:exitB1}
\end{equation}
\begin{equation}
\begin{split}
&\left.\frac{d\calE_\Csr}{ds}\right|_{D_1}= N r_c m c^2\sum_{n=0}^{\infty} (2-\delta_{n,0})(-1)^n  \int_0^{D_1} \dd L\left \{ \frac{T_L}{R_{L,n}^3}\lambda(s_{L,n})+\left[\beta^2 \frac{\Cos{\kappa_1 B_1}}{R_{L,n}}-\beta \frac{T_L}{R_{L,n}^2} \right]\lambda'(s_{L,n})\right\}\\
R_{L,n}&\equiv \sqrt{ \left( L+(s-B_1)\Cos{\kappa_1 B_1} +\frac{\Sin{\kappa_1 B_1}}{\kappa_1}\right)^2+\left(\frac{\Cos{\kappa_1 B_1}-1}{\kappa_1}-(s-B_1)\Sin{\kappa_1 B_1}\right)^2 +\left(n\,H\right)^2} \\
T_L &\equiv s-B_1 +L \Cos{\kappa_1 B_1}+\frac{1}{\kappa_1}\Sin{\kappa_1 B1}\\
s_{L,n} &\equiv -L-s_0 + \beta \sqrt{R_L^2+(n H)^2}.
\end{split}
\label{eq:exitD1}
\end{equation}
\end{widetext}


\end{document}